\documentclass[aps,prd, amsmath, amssymb, amsfonts,
floatfix,preprintnumbers]{revtex4-1}
\usepackage{graphicx}
\usepackage{multirow}
\usepackage{todonotes}
\usepackage{nicefrac}

\renewcommand{\vec}[1]{\mathbf{#1}}
\def\openone{\hbox{\upshape \small1\kern-3.3pt\normalsize1}}

\begin{document}

\title{The scalar pion form factor in two-flavor lattice QCD}

\author{Vera \surname{G\"ulpers}$^{1,2}$}
\author{Georg \surname{von Hippel}$^1$}
\author{Hartmut \surname{Wittig}$^{1,2}$}

\affiliation{$^1$PRISMA Cluster of Excellence and Institut f\"ur Kernphysik,
Johannes Gutenberg-Universit\"at Mainz, 55099 Mainz, Germany\\
$^2$Helmholtz Institute Mainz, Johannes Gutenberg-Universit\"at 
Mainz, 55099 Mainz, Germany}

\preprint{MITP/13-049}
\preprint{HIM-2013-04}

\begin{abstract}
We calculate the scalar form factor of the pion using two dynamical flavors of
non-perturbatively $\mathcal{O}(a)$-improved Wilson fermions, including both
the connected and the disconnected contribution to the relevant correlation
functions. We employ the
calculation of all-to-all propagators using stochastic sources and a
generalized hopping parameter expansion. From the form factor data at
vanishing momentum transfer, $Q^2=0$, and two non-vanishing $Q^2$
we obtain an
estimate for the scalar radius $\left<r^2\right>^\pi_{_{\rm S}}$ of the pion at
one value of the lattice spacing and for five different pion masses.  Using
Chiral Perturbation Theory at next-to-leading order, we find
$\left<r^2\right>^\pi_{_{\rm S}}=0.635\pm0.016\
\textnormal{fm}^2$ at the physical pion mass (statistical
error only). This is
in good agreement with the phenomenological estimate from $\pi\pi$-scattering.
The inclusion of the disconnected contribution is essential for achieving this
level of agreement.
\end{abstract}

\maketitle

\section{Introduction}

\def\xPT{\ensuremath{\chi}PT}

Recent years have seen extensive efforts to gain a quantitative understanding
of the low-energy dynamics of hadrons. The principal theoretical tools in
this endeavour are Chiral Perturbation Theory (\xPT)
\cite{Gasser:1983yg,Gasser:1984gg}
and numerical simulations of QCD on a space-time lattice. While \xPT\ is an
effective theory based on hadronic degrees of freedom, lattice QCD seeks to
describe hadronic properties from first principles in terms of the fundamental
constituents, i.e. the quarks and gluons. Lattice QCD and \xPT\ interact in
two ways: on the one hand, for performance reasons, lattice simulations are
usually performed at unphysically heavy light quark masses (although recently,
simulation results at physical light quark masses and below
\cite{Durr:2010aw,Durr:2010vn,Aoki:2009ix}
have become available), and thus \xPT\ is used to extrapolate results
obtained in a range of masses to the physical point, in order to obtain
physical predictions; on the other hand, lattice simulations allow for 
the calculation of low-energy matrix elements that can also be computed in
\xPT. Thus the low-energy constants of \xPT\ can be determined from first
principles
(cf. e.g. \cite{Heitger:2000ay,Giusti:2003iq,Giusti:2004yp,Gattringer:2005ij,
                Hasenfratz:2008ce,Beane:2011zm,Bernardoni:2011fx,
                Damgaard:2012gy,Borsanyi:2012zv,Herdoiza:2013sla}).
An important long-term goal is the quantitative description of nucleon
properties for which a wealth of data has been accumulated by numerous
experiments. However, baryonic systems are more difficult to treat
theoretically: while the range of validity of baryonic \xPT\ is largely
unknown, one finds that baryonic correlation functions computed in lattice QCD
suffer from an exponentially increasing noise-to-signal ratio. Therefore, the
interplay between lattice QCD and \xPT\ has mostly been studied in the context
of mesonic systems. In addition to investigations of masses and decay
constants, the focus has recently shifted to dynamical observables, such as
form factors, which depend on a momentum transfer. For instance, the vector
form factor, which describes the coupling of a photon to the pion and is thus
directly accessible to experiment, has been calculated to a fair level of
accuracy in lattice simulations
\cite{Capitani:2005ce,Brommel:2006ww,Jiang:2006gna,Kaneko:2007nf,
      Alexandrou:2007pn,Boyle:2008yd,Aoki:2009qn,Nguyen:2011ek,
      Fukaya:2012dla,Brandt:2013mb}.
While some of the systematics remain to be understood, the various
determinations of the pion charge radius, $\langle r^2\rangle^\pi_{_{\rm V}}$,
are mostly compatible with one another and also consistent with experiment.
On the other hand, the scalar pion form factor, defined by
\begin{equation}
  F^\pi_{_{\rm S}}\left(Q^2\right) \equiv
  \left<\pi^+\left(p_f\right)\right|\,m_{\rm d}\overline{d}d 
    +m_{\rm u}\overline{u}u\,\left|\pi^+\left(p_i\right)\right>,
    \qquad Q^2=-q^2=-(p_f-p_i)^2 
\label{eq:defff}
\end{equation}
is not directly accessible to experiment, since the Higgs (whose coupling
to the pion is determined by this form factor) is far too heavy to matter
in the low-energy regime of QCD. However, the scalar radius
\begin{equation}
 \left\langle r^2\right\rangle^\pi_{_{\rm S}} = - \frac{6}{F^\pi_{_{\rm
       S}}(0)} \frac{\partial F^\pi_{_{\rm S}}(Q^2)}{\partial Q^2}
 \Big|_{Q^2=0} \label{eq:scalarr} 
\end{equation}
of the pion can be related in \xPT\ to the ratio of the pion decay constant and
its value at vanishing quark mass via
\cite{Gasser:1983kx}
\begin{equation}
\frac{F_\pi}{F} = 1
                + \frac{1}{6} M_\pi^2 \left\langle r^2\right\rangle^\pi_{_{\rm
S}}
                + \frac{13M_\pi^2}{192\pi^2F_\pi^2} + O(M_\pi^4) \,.
\end{equation}
The scalar radius can also be linked to $\pi\pi$-scattering amplitudes
\cite{Donoghue:1990xh,Gasser:1990bv,Moussallam:1999aq},
and the most recent phenomenological estimate of ref.
\cite{Colangelo:2001df},
based on this approach, is $\left\langle
r^2\right\rangle^\pi_{_{\rm{S}}}=0.61\pm0.04$~fm$^2$.

The chiral expansion of the pion scalar radius at next-to-leading order (NLO)
\cite{Gasser:1983kx}
contains only a single low-energy constant $\bar\ell_4$. Since $\bar\ell_4$
also appears in the NLO expressions of other observables, one can test the
consistency of \xPT\ by comparing the lattice estimate of $\bar\ell_4$
extracted from the scalar form factor with that obtained from pseudoscalar
meson decay constants. Moreover, computing the pion scalar form factor in
lattice QCD gives a first-principles determination of $\bar\ell_4$ without any
modelling assumption, which would otherwise be implicit in a phenomenological
estimate. Another interesting feature of the pion scalar radius, from a
more technical point of view, is that a recent calculation in partially
quenched \xPT\
\cite{Juttner:2011ur,Juttner:2012xs}
indicates that the disconnected contribution to the scalar radius is
not negligible.

Determining the scalar form factor of the pion in lattice QCD is
computationally very demanding, due to the occurrence of quark-disconnected
diagrams (see figure \ref{fig:3ptdiagrams}). Such contributions are absent in
the corresponding hadronic matrix element of the vector current as a result
of charge conjugation invariance. Disconnected diagrams are expensive to
compute on the lattice, because they require the trace of the propagator from
a point to itself to be evaluated; in order to reliably estimate this
quantity, it is necessary to compute the propagator from each point of the
lattice to itself. Naively, this would require an inversion of the lattice
Dirac operator for each lattice point, which is prohibitively
expensive. Efficient methods to calculate such all-to-all propagators have
therefore been developed, including the use of noisy sources
\cite{Bitar:1988bb},
low-mode averaging
\cite{Neff:2001zr,Giusti:2004yp,Bali:2005fu},
hopping parameter expansions
\cite{Thron:1997iy},
and truncated solver methods
\cite{Collins:2007mh}.
Nevertheless, the computational effort involved is significant.
The pion scalar form factor is therefore far less well studied than the
vector form factor; so far only one calculation of the full scalar form
factor
\cite{Aoki:2009qn},
which has been performed on a rather small $32\times16^3$ lattice, exists.

In this paper we expand on our account in
\cite{Gulpers:2012kd}
by presenting the details and results of our calculation of the pion
scalar form factor using $\mathcal{O}(a)$-improved Wilson fermions.
Details of the lattice ensembles and observables used are given in
section~\ref{sec:sim}, and the methods used to calculate the disconnected
contribution using a combination of stochastic sources and a generalized
hopping parameter expansion are described in section~\ref{sec:inv}.
Our data analysis methods are detailed in section~\ref{sec:ratio}, and the
results for the form factor, as well as the scalar radius, including the
determination of the low-energy constant $\bar\ell_4$ from the chiral
extrapolation of the scalar radius are given in section~\ref{sec:results}.
We conclude with a summary of our main findings and
several remarks on the differences between our results and those of
\cite{Aoki:2009qn}
in section~\ref{sec:conclusions}.

\section{Simulation Setup}
\label{sec:sim}

Our calculation of the scalar pion form factor is performed with
$N_f=2$ dynamical flavors of non-perturbatively $\mathcal{O}(a)$-improved
Wilson fermions. The corresponding Dirac operator $D_{_{SW}}$ is given by
\begin{equation}
D_{\rm{sw}} = D_{\rm{w}} + c_{\rm{sw}}\,\frac{i}{4} \sigma_{\mu\nu}
\hat{F}_{\mu\nu} \label{eq:SW}
\end{equation}
where 
\begin{equation}
  D_{\rm{w}}=\frac{1}{2\kappa}\,\openone -\frac{1}{2}\,H\label{eq:WilsonDirac}
\end{equation}
is the unimproved Wilson-Dirac operator, and the term with coefficient
$c_{\rm{sw}}$ in \eqref{eq:SW} is the Sheikholeslami-Wohlert (clover) term
\cite{Sheikholeslami:1985ij}
implementing $\mathcal{O}(a)$-improvement
\cite{Luscher:1996sc}.
Since the latter is local, all couplings between neighboring lattice points
appearing in \eqref{eq:WilsonDirac} are contained in the hopping matrix $H$.
The hopping parameter $\kappa$ determines the bare quark mass
\begin{equation}
m = \frac{1}{2a}\left(\frac{1}{\kappa}-\frac{1}{\kappa_c}\right)\,,
\end{equation}
where $\kappa_c$ is the critical value for which the quark (and hence pion)
mass vanishes. For our simulations we use gauge ensembles produced as part of
the CLS initiative, which have been generated using L\"uscher's
deflation-accelerated DD-HMC
algorithm 
\cite{Luscher:2005rx,Luscher:2007es}.
An overview of the ensembles used in this study can be found in
table~\ref{tab:ensembles}. Here we use the non-perturbative determination of
the improvement coefficient $c_{\rm{sw}}$ for $N_f=2$ flavors
\cite{Jansen:1998mx}
at a single value of the gauge coupling, $\beta=5.3$. The corresponding
lattice spacing of $a=0.063$~fm was determined via the mass of the $\Omega$
baryon 
\cite{Capitani:2011fg}.
A similar result for the lattice spacing was obtained by the ALPHA
collaboration using the Kaon decay constant
\cite{Fritzsch:2012wq}.
\begin{table}[h]
 \begin{tabular}{ccccccccccccccccc}
\hline\hline
$\beta$ &&  $a [\textnormal{fm}]$ && lattice && $m_\pi [\textnormal{MeV}]$ &&
$m_\pi L$ && $\kappa$ && Label && $N_{\rm{cfg}}$\\
\hline
$5.3$ && $0.063$  && $64\times32^3$ && 650 && 6.6 && $0.13605$ && E3 && $156$\\
$5.3$ && $0.063$  && $64\times32^3$ && 605 && 6.2 && $0.13610$ && E4 && $162$\\
$5.3$ && $0.063$  && $64\times32^3$ && 455 && 4.7 && $0.13625$ && E5 && $1000$\\
\hline
$5.3$ && $0.063$  && $96\times48^3$ && 325 && 5.0 && $0.13635$ && F6 && $300$\\
$5.3$ && $0.063$  && $96\times48^3$ && 280 && 4.3 && $0.13638$ && F7 && $351$\\
\hline\hline
 \end{tabular}
\caption{Overview of the CLS ensembles used in this work. The lattice spacing
         given was determined using the $\Omega$ baryon mass
         \cite{Capitani:2011fg}.
         Note that all ensembles fulfill $m_\pi L>4$.
         }
\label{tab:ensembles}
\end{table}

\section{Calculation of disconnected diagrams}
\label{sec:inv}

\subsection{Inversion with stochastic sources}
\label{subsec:stochsources}

\begin{figure}[t]
\centering
\includegraphics[trim = 25mm 238mm 10mm 33mm, scale=0.75]{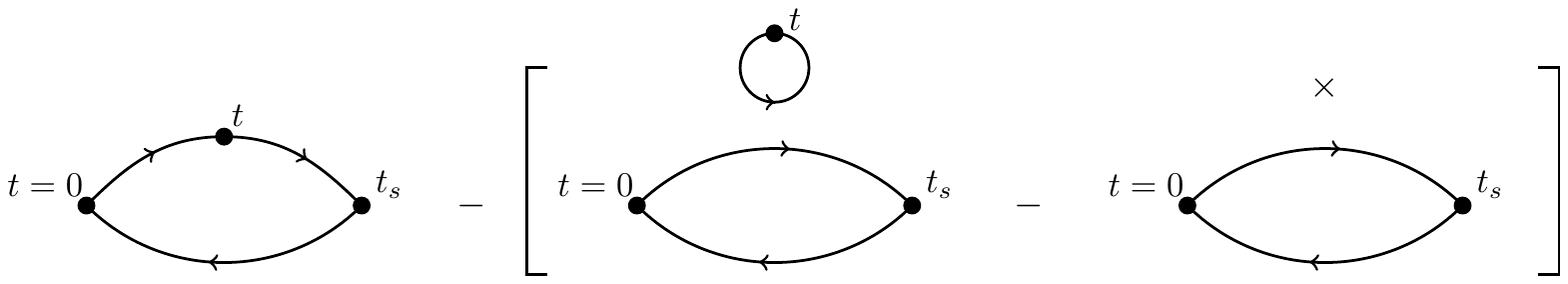}
\caption{The three contributions to the three-point function. The connected
on the left, the disconnected with
subtracted vacuum on the right. The middle diagram
contains the loop factor $L(\vec{p},t)$.}\label{fig:3ptdiagrams}
\end{figure}

While the connected three-point function can be calculated using conventional
point-to-all propagators and the extended propagator method 
\cite{Martinelli:1988rr},
the disconnected three-point function is computationally more demanding,
since the calculation of the loop $L(\vec{p},t)$
(c.f. figure \ref{fig:3ptdiagrams}) requires the all-to-all propagator,
i.e. the inverse of a generic lattice Dirac operator~$D$ for arbitrary source
and sink positions:
\begin{equation}
L(\vec{p},t) = \sum\limits_{\vec{x}}
e^{i\vec{p}\cdot\vec{x}}\,\,\textnormal{Tr}\left[\Gamma
  D^{-1}(x,x)\right]\,.\label{eq:loop} 
\end{equation}
One particular method for calculating the all-to-all propagator is based on
the use of stochastic sources
\cite{Bitar:1988bb,Bali:2009hu}.
As a first step one selects $N$ random source vectors, $\left|\eta_i\right>$,
which fulfill the conditions
\begin{equation}
\frac{1}{N}\sum\limits_{i=1}^N \left|\eta_i\right> =
0 +
\mathcal{O}\left(1/\sqrt{N}\right)\hspace{0.3cm}\textnormal{,}\hspace{2.5cm}
\frac{1}{N}\sum\limits_{i=1}^N
\left|\eta_i\right>\left<\eta_i\right|=\openone +
\mathcal{O}\left(1/\sqrt{N}\right)\,. \label{eq:condstochsources}
\end{equation}
After solving the Dirac equation
$D\,\left|s_i\right> = \left|\eta_i\right>$ for all $N$ sources,
an estimate of the propagator is given by
\begin{equation}
  D^{-1} = \frac{1}{N} \sum\limits_{i=1}^N
\left|s_i\right>\left<\eta_i\right|\,.\label{eq:invwithstoch}
\end{equation}
While the statistical error associated with the stochastic noise scales like
$N^{-1/2}$, the numerical cost of the method is proportional to the number of
stochastic sources, $N$. It is then clear that one has to optimize the value
of $N$, in order to balance good statistical accuracy against an acceptable
numerical effort. The generalized hopping parameter expansion described in the
following section is designed to reduce the statistical error of the
disconnected contribution for a given number of stochastic sources.

\subsection{The generalized Hopping Parameter Expansion}

The inverse of the Wilson-Dirac operator can be expressed in terms of a
hopping parameter expansion (HPE)
\cite{Thron:1997iy,Bali:2009hu}. 
As already indicated in \eqref{eq:WilsonDirac}, the unimproved Wilson-Dirac
operator can be split into two parts, one of which is proportional to the unit
matrix while the other matrix, the hopping term $H$, contains all couplings of
neighboring lattice points, 
\begin{equation}
 D_{\rm{w}}=\frac{1}{2\kappa}\,\openone -\frac{1}{2}\,H\,,
\end{equation}
where $\kappa$ denotes the hopping parameter. For the calculation of the quark
propagator $D_{\rm{w}}^{-1}$, the hopping parameter expansion amounts to
performing a geometric series expansion in $\kappa$,
\begin{align}
 D_{_{W}}^{-1}& = 2\kappa \sum\limits_{i=0}^{k-1} \left(\kappa\,H\right)^{i} +
\left(\kappa\,H\right)^{k} D_{_{W}}^{-1} \,. \label{eq:hpewoimp}
\end{align}
The advantage of rewriting the propagator in this way lies in the fact that
$D_{\rm{w}}^{-1}$ on the right-hand side is multiplied by $k$ powers of
$\kappa<1$. Hence one expects that the noise introduced by the stochastic
inversion of $D_{\rm{w}}$ is reduced accordingly.

When $\mathcal{O}(a)$-improvement is employed, equation \eqref{eq:hpewoimp}
must be generalized. According to equation \eqref{eq:SW} the improved operator
has
the form 
\begin{equation}
  D_{\rm{sw}}=\frac{1}{2\kappa}\,\openone -\frac{1}{2}\,H + c_{\rm{sw}} B
\,,\label{eq:WD}
\end{equation}
where $B=\frac{1}{4}\sigma_{\mu\nu}F_{\mu\nu}$ is the clover term. This can be
rewritten as 
\begin{equation}
 D_{_{SW}} =
A-\frac{1}{2}\,H=A\left(\openone-\frac{1}{2}\,A^{-1}H\right)\hspace{0.5cm}
\textnormal{where}\hspace{0.3cm}A =
\frac{1}{2\kappa}\,\openone + c_{_{SW}} B\,,\label{eq:Drewritten}
\end{equation}
which again allows for a geometric series expansion, resulting in
\begin{equation}
 D_{_{SW}}^{-1} = \sum\limits_{i=0}^{k-1}
\left(\frac{1}{2}\,A^{-1}\,H\right)^{i}\,A^{-1} +
\left(\frac{1}{2}\,A^{-1}\,H\right)^{k} D_{_{SW}}^{-1}\,.\label{eq:hpe}
\end{equation}
In \eqref{eq:hpe}, the inverse of the matrix $A$, which is defined
in \eqref{eq:Drewritten}, appears. Without $\mathcal{O}(a)$-improvement, i.e.
$c_{_{SW}}=0$, this inverse is trivial, $A^{-1}=2\kappa$, and \eqref{eq:hpe}
reduces to \eqref{eq:hpewoimp}. For $c_{_{SW}}\neq0$, one can show that the
matrix $A$ is block-diagonal due to the local form of the clover term.
Therefore one only has to invert two $6\times6$ matrices for each lattice
point, which is still comparatively cheap in terms of the required computer
time.

The inverse $D_{^{SW}}^{-1}$ on the right-hand side of \eqref{eq:hpe} can now
be estimated with stochastic sources as described above. In order to find a
good compromise between statistical fluctuations and low numerical cost, one
can now tune two parameters, namely the number of stochastic sources $N$ and
the order $k$ of the hopping parameter expansion.

\begin{figure}[h]
 \centering
\includegraphics[scale=0.65]{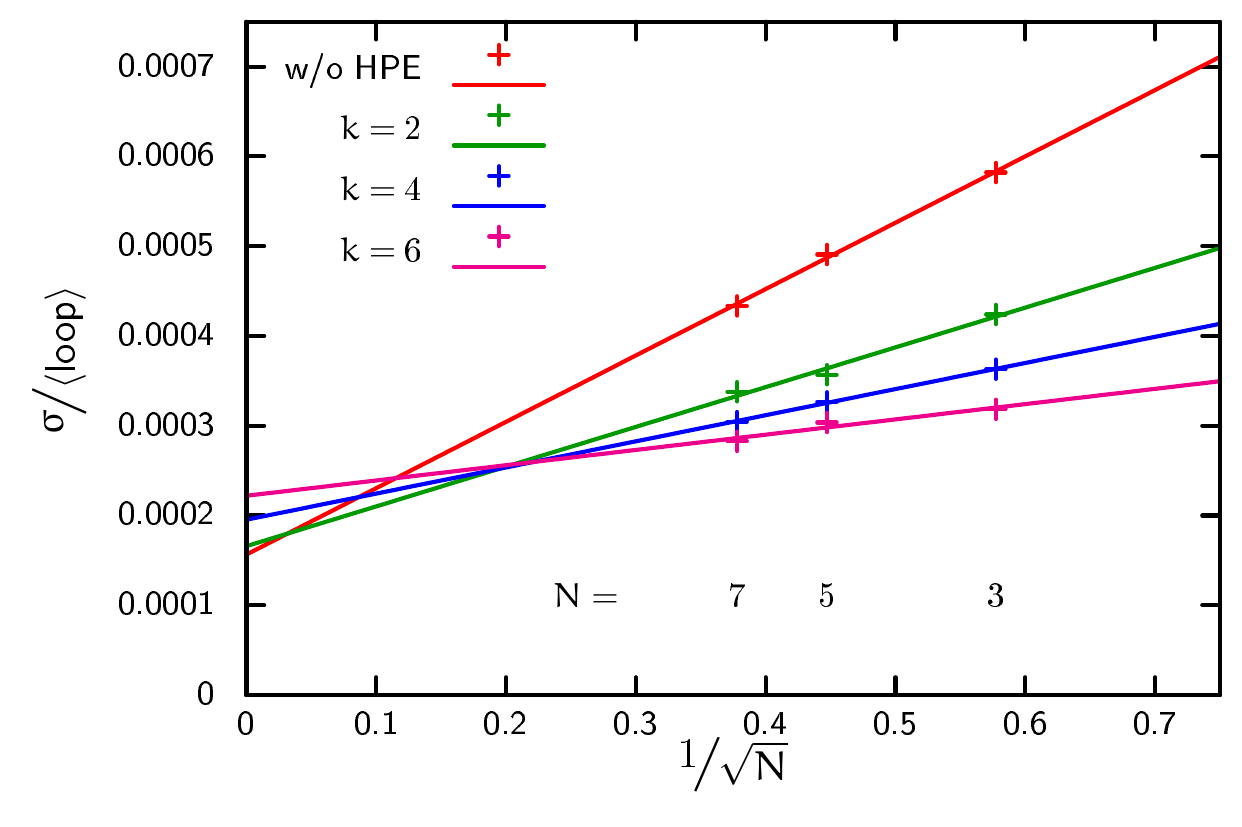}
\caption{The relative statistical error of the loop $L(\vec{p}=0,\,t=0)$}
\label{fig:loopgaugemean}
\end{figure}

As an example how these two parameters influence the effort required to reach
a given statistical precision, we show in figure \ref{fig:loopgaugemean} the
standard deviation of the loop $L(\vec{p}=0,\,t=0)$ divided by its gauge mean
(i.e. the relative statistical error) computed on 33
configurations of the E4 ensemble (cf. table \ref{tab:ensembles}). The loop
has been calculated stochastically without employing the HPE, as well as for
$k=2$, $4$, $6$ terms in the hopping parameter expansion, using $N=3$, $5$ and
$7$ sources in each case. One can see clearly that increasing the order of the
HPE decreases the statistical error of the loop. In addition, we observe the
expected behavior for the scaling of the error, $\sigma\propto\sqrt{N}^{-1}$,
as indicated by the linear curves in figure \ref{fig:loopgaugemean}. Therefore
the intercept on the $y$-axis shows the remaining gauge noise in the
calculation. To obtain a good balance between the accuracy of the calculation
and the computer time needed, we use $N=3$ stochastic sources and the order
$k=6$ of the generalized HPE for the calculation of the loop. At this point
the error is already close to the gauge noise, and the relatively small gain
in statistical accuracy does not justify a further increase in the number of
stochastic sources $N$.

In order that the method produces an exact result for the loop, also the
contributions from the first $k$ terms in the generalized hopping parameter
expansion of equation~\eqref{eq:hpe}, i.e.
\begin{equation}
X\equiv\,\, \sum\limits_{i=0}^{k-1}
\mathrm{tr~}\left[\Gamma\left(\frac{1}{2}\,A^{-1}\,H\right)^{i}\,A^{-1}\right]\,
, \label{eq:otherterms}
\end{equation}
have to be calculated. This can also be done with stochastic sources, by
inserting a unit matrix in \eqref{eq:otherterms} and using
\eqref{eq:condstochsources}, i.e.
\begin{equation}
\mathrm{tr~}X = \mathrm{tr~}(X\openone) = \frac{1}{M}\sum\limits_{i=1}^M
\mathrm{tr~}\left(X\left|\eta_i\right>\left<\eta_i\right|\right) +
\mathcal{O}\left(1/\sqrt{M}\right)
 = \frac{1}{M}\sum\limits_{i=1}^M
\left<\eta_i\right|X\left|\eta_i\right> + \mathcal{O}\left(1/\sqrt{M}\right)\,.
\end{equation}
Since this calculation does not require much computer time compared to the
inversion, we can use a large number $M=50$ of sources. A more detailed
discussion of the tuning of the generalized hopping parameter expansion
can be found in
\cite{Gulpers:Diplom}.

\section{Extracting the form factor}
\label{sec:ratio}

\subsection{Two- and three-point functions}

The scalar form of the pion can be determined from appropriate combinations of
the two- and three-point correlation functions. In order to compute the ground
state energy of a pion with momentum $\vec{p}$ we consider the two-point
function
\begin{equation}
C_{2\textrm{pt}}(t,\vec{p}) = \sum_{\vec{x}} e^{-i\vec{p}\cdot\vec{x}} 
                                \langle \phi(t,\vec{x})\phi(0)\rangle
\end{equation}
of the pseudoscalar density
\begin{equation}
\phi(x)=\overline{q}(x)\gamma_5 q(x)\,.
\label{eq:piop}
\end{equation}
On a periodic lattice with time extent~$T$ the asymptotic behavior at large
Euclidean times~$t$ is given by
\begin{equation}
C_{2\textnormal{pt}}(t,\vec{p})  \sim
\frac{Z(\vec{p})^2}{2 E_\pi(\vec{p})}\left[e^{- t E_\pi(\vec{p})}
+ e^{-(T-t)E_\pi(\vec{p})} \right]\,,
\label{eq:2pt}
\end{equation}
where $E_\pi(\vec{p})$ is the energy of the pion, and
$Z(\vec{p})^2=\left|\left<\pi(\vec{p})\right|\phi(0)\left|0\right>\right|^2$
is the squared matrix element of the pseudoscalar density between a pion state
and the vacuum.

In order to describe the coupling of a scalar particle to the pion, one has to
consider insertions of the local scalar density
\begin{equation}
\mathcal{O}_{\rm S}(y)=\overline{q}(y)q(y)\,.
\end{equation} 
The scalar form factor can be extracted from the three-point correlation
function
\begin{equation}
 C_{3\textnormal{pt}}(t,t_s,\vec{p}_i,\vec{p}_f) = 
    \sum_{\vec{x},\vec{y}} e^{-i\vec{p}_f\cdot\vec{x}+i\vec{q}\cdot\vec{y}}
     \langle\phi(t_s,\vec{x})\mathcal{O}(t,\vec{y})\phi(0)\rangle\,,
\end{equation}
where $\vec{p}_i, \vec{p}_f$ denote the three-momenta of the initial and final
pions, respectively, and $Q^2=-q^2=-(p_f-p_i)^2$ is the squared momentum
transfer. For $0\ll t\ll t_s$ the three-point functions behaves like
\begin{equation}
  C_{3\textnormal{pt}}(t,t_s,\vec{p}_i,\vec{p}_f) \sim
    \frac{Z(\vec{p}_i) Z(\vec{p}_f)}{4E_\pi(\vec{p}_i)E_\pi(\vec{p}_f)}
\left<\pi(\vec{p}_f)\right|\mathcal{O}_{\rm{S}}(0)\left|\pi(\vec{p}_i)\right>
    e^{-(t_s-t)E_\pi(\vec{p}_f)}e^{-t E_\pi(\vec{p}_i)}\,, \label{eq:3pt}
\end{equation}
and the matrix element
$\left<\pi(\vec{p}_f)\right|\mathcal{O}_{\rm{S}}(0)\left|\pi(\vec{p}_i)\right>$
that occurs in equation\,\eqref{eq:3pt} is the desired scalar form
factor. Note that in the scalar case the vacuum contribution
\begin{equation}
C_{\textrm{vac}}(t,t_s,\vec{p}_i,\vec{p}_f) = C_{2\textrm{pt}}(t_s,\vec{p}_f) 
  \sum_{\vec{y}} e^{i\vec{q}\cdot\vec{y}}\left<\mathcal{O}_{\rm{S}}(t,\vec{y})
  \right>
\end{equation}
is non-zero for $\vec{q}=0$ and must be subtracted prior to fitting numerical
data for $C_{3\textnormal{pt}}$ to
equation~\eqref{eq:3pt}. Figure\,\ref{fig:3ptdiagrams} shows the three
diagrams that contribute to the three-point function, i.e. the quark-connected
and disconnected diagrams, as well as the subtracted vacuum contribution.

Our simulations are performed using Wilson fermions, which break chiral
symmetry explicitly. As a consequence, the scalar operator
$\mathcal{O}=\overline{q}q$ undergoes an additive renormalization besides the
multiplicative one, i.e.
\begin{equation}
 \left<\mathcal{O}^{\rm R}_{\rm S}\right> = Z_{{\rm S}}
 \left<\mathcal{O}_{\rm S}-b_0\right>\,.
\end{equation}
The subtraction of the vacuum contribution (cf. figure \ref{fig:3ptdiagrams})
ensures that the cubically divergent additive renormalization $b_0$ of the
scalar operator is canceled. Since the multiplicative renormalization constant
$Z_{\rm{S}}$ has not been determined in our calculation, all form factor data in
this paper are not renormalized. Note, however, that $Z_{\rm{S}}$ drops out in
the calculation of the scalar radius (cf. equation \eqref{eq:scalarr}), which
implies that our results can be readily compared to phenomenology and other
lattice determinations.

\subsection{Building Ratios}

To extract the scalar matrix element
$\left<\pi(\vec{p}_f)\right|\mathcal{O}_{\rm{S}}(0)\left|\pi(\vec{p}_i)\right>$,
it is convenient to form appropriate ratios of three- and two-point
functions. Here we follow the approach of ref.
\cite{Boyle:2007wg},
focusing, in particular, on the two ratios called $R_1$ and $R_3$,
\begin{align}
  &R_1(t,t_s,\vec{p}_i,\vec{p}_f) = 
\sqrt{\frac{C_{3\textnormal{pt}}(t,t_s,\vec{p}_i,\vec{p}_f)
 C_{3\textnormal{pt}}(t,t_s,\vec{p}_f,\vec{p}_i)}
{C_{2\textnormal{pt}}(t_s,\vec{p}_i)C_{2\textnormal{pt}}(t_s,\vec{p}_f)}}\,\,,
\label{eq:Ratio1}\\
 &R_3(t,t_s,\vec{p}_i,\vec{p}_f) =
\frac{C_{3\textnormal{pt}}(t,t_s,\vec{p}_i,\vec{p}_f)}
{C_{2\textnormal{pt}}(t_s,\vec{p}_f)}\cdot\sqrt{\frac{C_{2\textnormal{pt}} 
(t_s, \vec{p}_f)
C_{2\textnormal{pt}}(t,\vec{p}_f)C_{2\textnormal{pt}}((t_s-t),\vec{p}_i)}
{C_{2\textnormal{pt}}(t_s,\vec{p}_i)C_{2\textnormal{pt}}(t,\vec{p}_i)
C_{2\textnormal{pt}}((t_s-t),\vec{p}_f) } }\label{eq:Ratio3}\,.
\end{align}
When the expressions of equations \eqref{eq:2pt} and \eqref{eq:3pt} for the
asymptotic forms of the two- and three-point functions are inserted into the
definition of $R_1$ one obtains
\begin{equation}
R_1(t,t_s,\vec{p}_i,\vec{p}_f)
\sim\frac{\left<\pi(\vec{p}_f)\right|\mathcal{O}_{\rm{S}}(0)
\left|\pi(\vec{p}_i)\right>}{2\sqrt{E_\pi(\vec{p}_i)E_\pi(\vec{p}_f)}}
\sqrt{\frac{e^{-E_\pi(\vec{p}_i) t_s}e^{-E_\pi(\vec{p}_f) t_s}}
{(e^{-E_\pi(\vec{p}_i) t_s} +
e^{-E_\pi(\vec{p}_i)(T-t_s)})\cdot(e^{-E_\pi(\vec{p}_f)
 t_s} + e^{-E_\pi(\vec{p}_f) (T-t_s)})}} \label{eq:R1}\,.
\end{equation}
Here all overlap factors $Z(\vec{p})$, as well as any dependence on the time
$t$ of the operator insertion cancel. The remaining dependence on the
source-sink separation $t_s$ is due to the backward propagating pion, and the
corresponding expression under the square root in equation \eqref{eq:R1}
approaches unity as $T\to\infty$. For any finite value of $T$, it is easily
determined, since all pion energies $E_\pi(\vec{p})$ are known from the
two-point functions.

Inserting equations \eqref{eq:2pt} and \eqref{eq:3pt} into the expression for
$R_3$ leads to
\begin{equation}
 R_3(t,t_s,\vec{p}_i,\vec{p}_f)
\sim\frac{\left<\pi(\vec{p}_f)\right|\mathcal{O}_{\rm{S}}(0)
\left|\pi(\vec{p}_i)\right>}{2\sqrt{E_\pi(\vec{p}_i)E_\pi(\vec{p}_f)}}\,f(t,
t_s)\, ,
\label{eq:R3}
\end{equation}
where the factor 
\begin{equation}
\begin{split}
f(t,t_s) = &\frac{e^{-(t_s-t)E_\pi(\vec{p}_f) E_\pi(\vec{p}_i)}}{(e^{-t_s
E_\pi(\vec{p}_f)} + e^{-(T-t_s) E_\pi(\vec{p}_f)})}\times\\
&\sqrt{\frac{(e^{-t_s E_\pi(\vec{p}_f)} + e^{-(T-t_s) E_\pi(\vec{p}_f)})(e^{-t
E_\pi(\vec{p}_f)} + e^{-(T-t)
E_\pi(\vec{p}_f)})(e^{-(t_s-t) E_\pi(\vec{p}_i)} + e^{-(T-(t_s-t))
E_\pi(\vec{p}_i)})}{(e^{-t_s E_\pi(\vec{p}_i)} + e^{-(T-t_s)
E_\pi(\vec{p}_i)})(e^{-t E_\pi(\vec{p}_i)} + e^{-(T-t)
E_\pi(\vec{p}_i)})(e^{-(t_s-t) E_\pi(\vec{p}_f)} + e^{-(T-(t_s-t))
E_\pi(\vec{p}_f)})}}
\end{split}
\end{equation}
depends on both $t_s$ and the time $t$ of the
operator insertion. As in the case of $R_1$, the time dependence can be
determined for every $t$ and $t_s$ once the pion energies are known from the
two-point functions. For large time separations $0\ll t\ll t_s\ll T/2$, the
factor $f(t,t_s)\rightarrow 1$, i.e. the ratio $R_3$ forms a plateau, which is
proportional to the form factor.

Note that equation \eqref{eq:R3} is only valid when the same interpolating
operator (e.g. with smeared or point-like quark fields) is used at the pion
source and the pion sink. Otherwise, not all factors $Z(\vec{p})$ cancel out,
since
they depend on the source type
\cite{Bonnet:2004fr}.
Moreover, the three-point functions must obviously be computed with the same
type of source and sink as the two-point functions.

In the calculation of the quark-connected contribution to the three-point
function, Gaussian smearing
\cite{Gusken:1989ad,Alexandrou:1990dq,Allton:1993wc}
was only applied at the source. Therefore, the
connected
part could only be determined via the ratio $R_1$. By contrast, for the
quark-disconnected part we had smeared-smeared pion two-point functions at our
disposal. Since we found that the ratio $R_3$ gives a much cleaner signal than
$R_1$, we have computed $R_3$ for smeared-smeared correlation functions, in
order to determine the quark-disconnected contribution.

\section{Results}
\label{sec:results}

In this section we present our results for the ratios from which the scalar
form factor can be determined. For these results to be reliable it is
important to address the issue of unwanted contributions from excited states
which may arise if the separations in Euclidean time are not large enough to
guarantee that the correlation functions $C_{2\textrm{pt}}$ and
$C_{3\textnormal{pt}}$ can be described by their asymptotic behavior. We have
therefore performed a systematic study of the $t_s$-dependence of the ratios
$R_1$ and $R_3$.

Twisted boundary conditions
\cite{Bedaque:2004kc,Sachrajda:2004mi,Flynn:2005in,deDivitiis:2004kq,
Boyle:2007wg}
are widely used to compute vector form factors for nearly arbitrary momentum
transfers $Q^2$. In the case of the scalar form this is not an option, since
the effect of the twist angle cancels in the quark-disconnected contribution. 
Therefore we discuss our results for vanishing momentum transfer, as well as
two non-zero values of $Q^2$ which can be realized via the usual
Fourier momenta.

\subsection{Ratios for $Q^2=0$}

In the case of vanishing momentum transfer, $Q^2=0$, i.e. for
$\vec{p}_i=\vec{p}_f=\vec{p}$, the ratios $R_1$ and $R_3$ are
identical. Specifically, for $\vec{p}_i=\vec{p}_f=0$ we have
\begin{equation}
R_1(t,t_s,0,0) \equiv R_3(t,t_s,0,0) = 
\frac{C_{3\textnormal{pt}}(t,t_s,0,0)}{C_{2\textnormal{pt}}
(t_s,0)}
\sim \frac{\left<\pi(0)\right|\mathcal{O}_{\rm{S}}(0)
\left|\pi(0)\right>}{2m_\pi}
\underbrace{\frac{e^{-m_\pi t_s}}{e^{-m_\pi t_s} +
e^{-m_\pi(T-t_s)}}}_{{=f(t_s)}}\,,\label{eq:Rq20}
\end{equation}
where we have assumed that the ground state
dominates. Equation~\eqref{eq:Rq20} can be used to extract the form factor for
$Q^2=0$ from the simulated three- and two-point function data at zero
momentum. To increase the statistics we have exploited translational invariance
by computing the disconnected contribution for four different pion source
positions separated by $T/4$.

\begin{figure}
\centering
\includegraphics[scale=0.65]{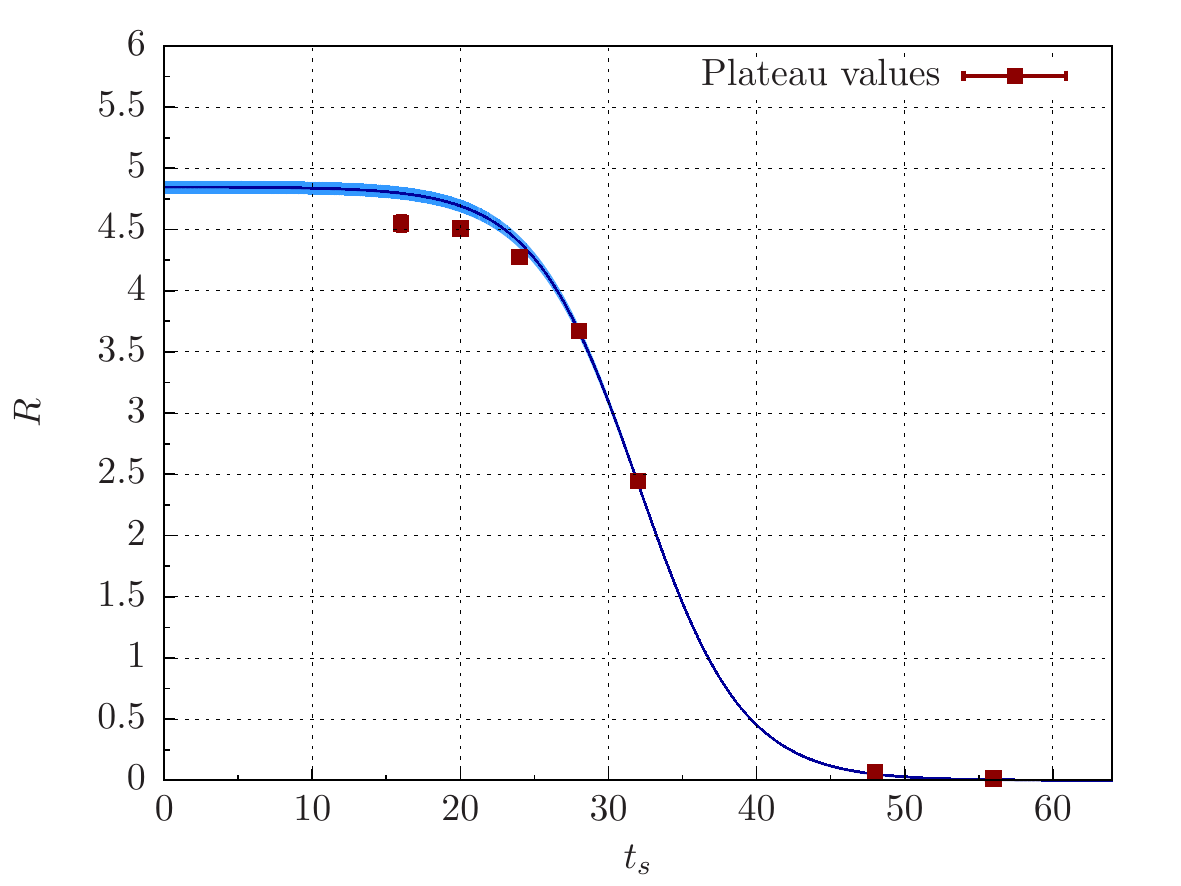} 
\includegraphics[scale=0.65]{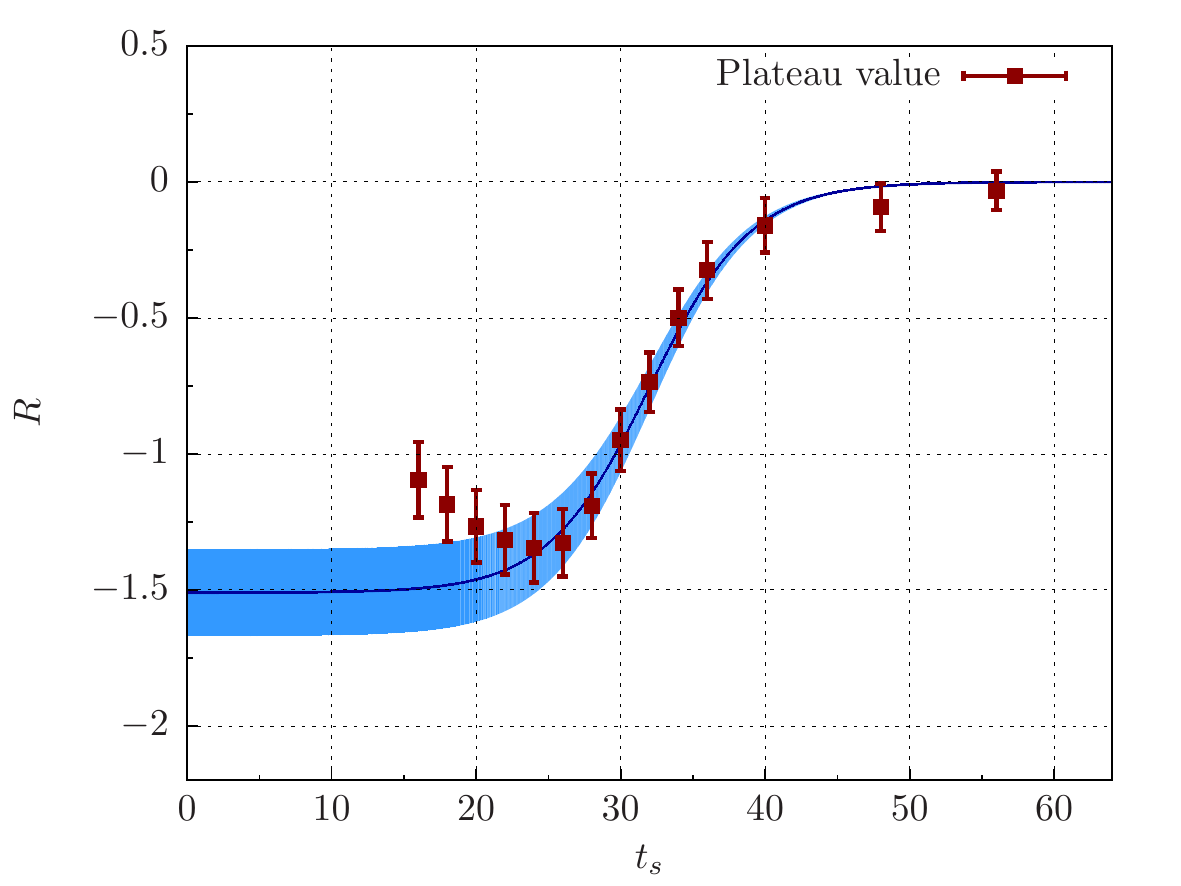}
\caption{Plateau values plotted against the different $t_s$ for
  vanishing momentum transfer $Q^2=0$ for the E5 ensemble. The
connected contribution
  (smeared-local) is shown on the left, the disconnected (smeared-smeared)
  on the right. A function of the form \eqref{eq:Rq20} has been
fitted to the data.}
\label{fig:mom0ratiovsts}
\end{figure}

To investigate the $t_s$-dependence of the ratios, we fitted constants
to the plateau regions of the ratios for the different values of $t_s$. The
plateau values obtained are plotted against $t_s$ in figure
\ref{fig:mom0ratiovsts} for the E5 ensemble, which has the highest statistics of
all ensembles studied so far. The blue line indicates a function of the form
\eqref{eq:Rq20}, which has been fitted to the data.  
Clearly, the data deviate from the expected $t_s$-dependence for the
smaller values $t_s<24$ for both the connected and the disconnected
contribution. However, for larger source-sink separations our data show the
expected $t_s$-dependence. The deviation at small $t_s$ indicates the presence
of excited state contributions for $t_s<24$.

\begin{figure}
\centering
\includegraphics[scale=0.65]{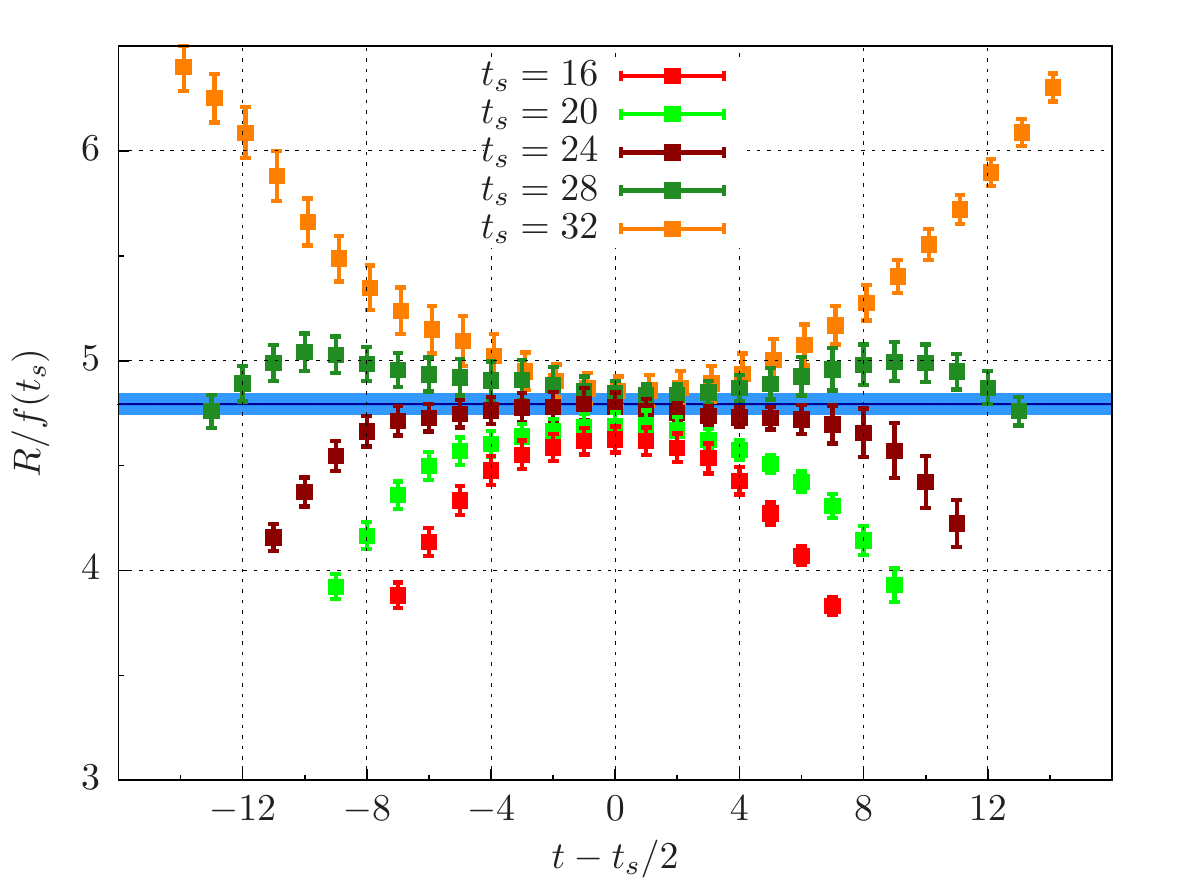} 
\includegraphics[scale=0.65]{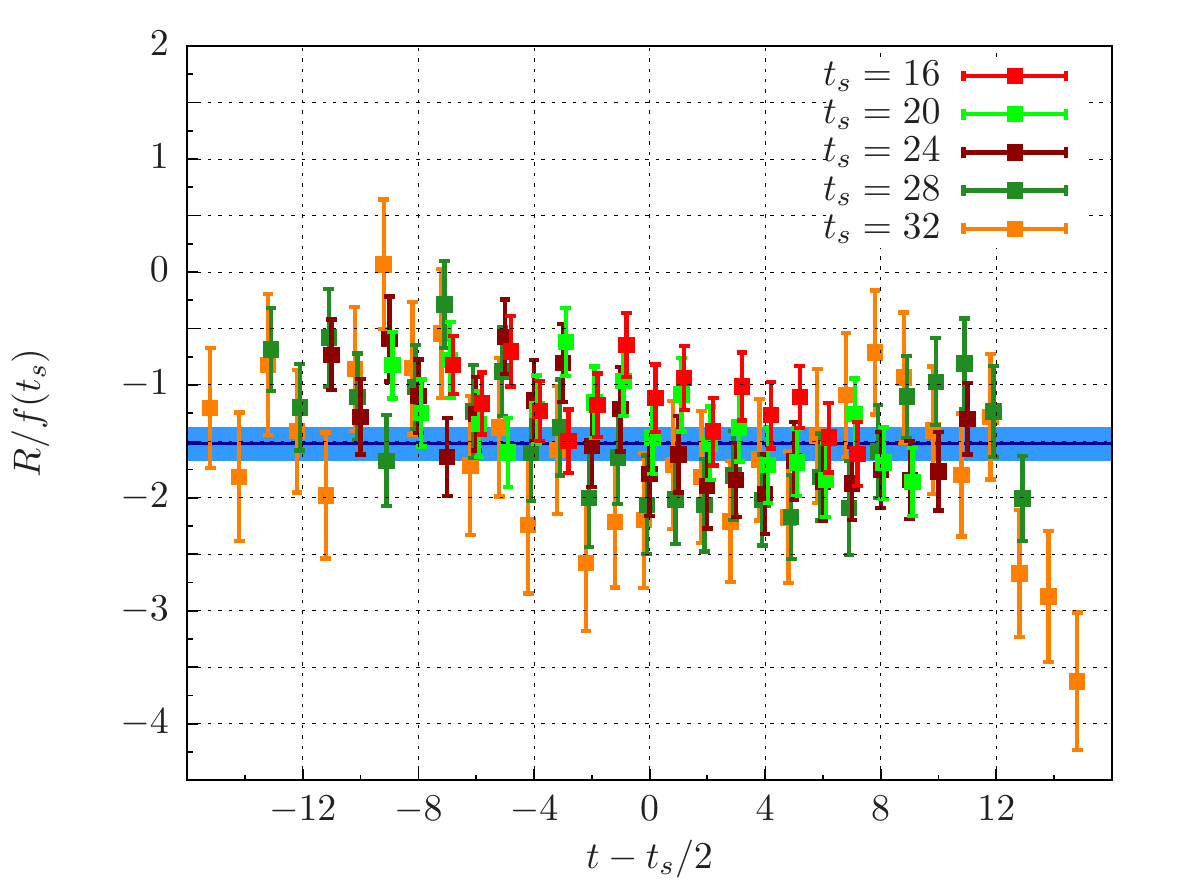}
\caption{Results for the ratios corrected by the $t_s$-dependence for
  vanishing momentum transfer $Q^2=0$ for the E5 ensemble. The
connected contribution
  (smeared-local) is shown on the left, the disconnected (smeared-smeared)
  on the right. The blue lines indicate the results
  of the global fit to a constant. The fit ranges in $t$ are listed in table
\ref{tab:fitrangest}.}
\label{fig:mom0}
\end{figure}
\begin{figure}
\centering
\includegraphics[scale=0.65]{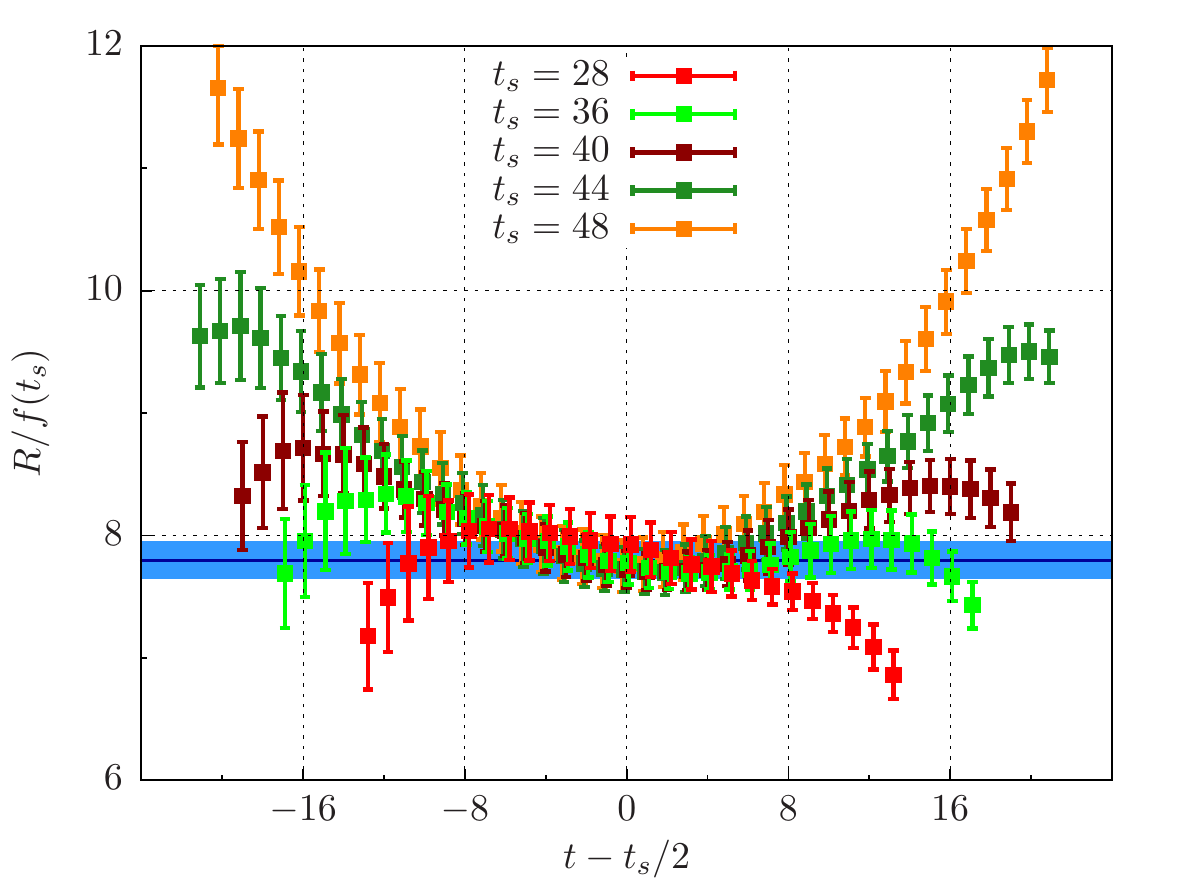} 
\includegraphics[scale=0.65]{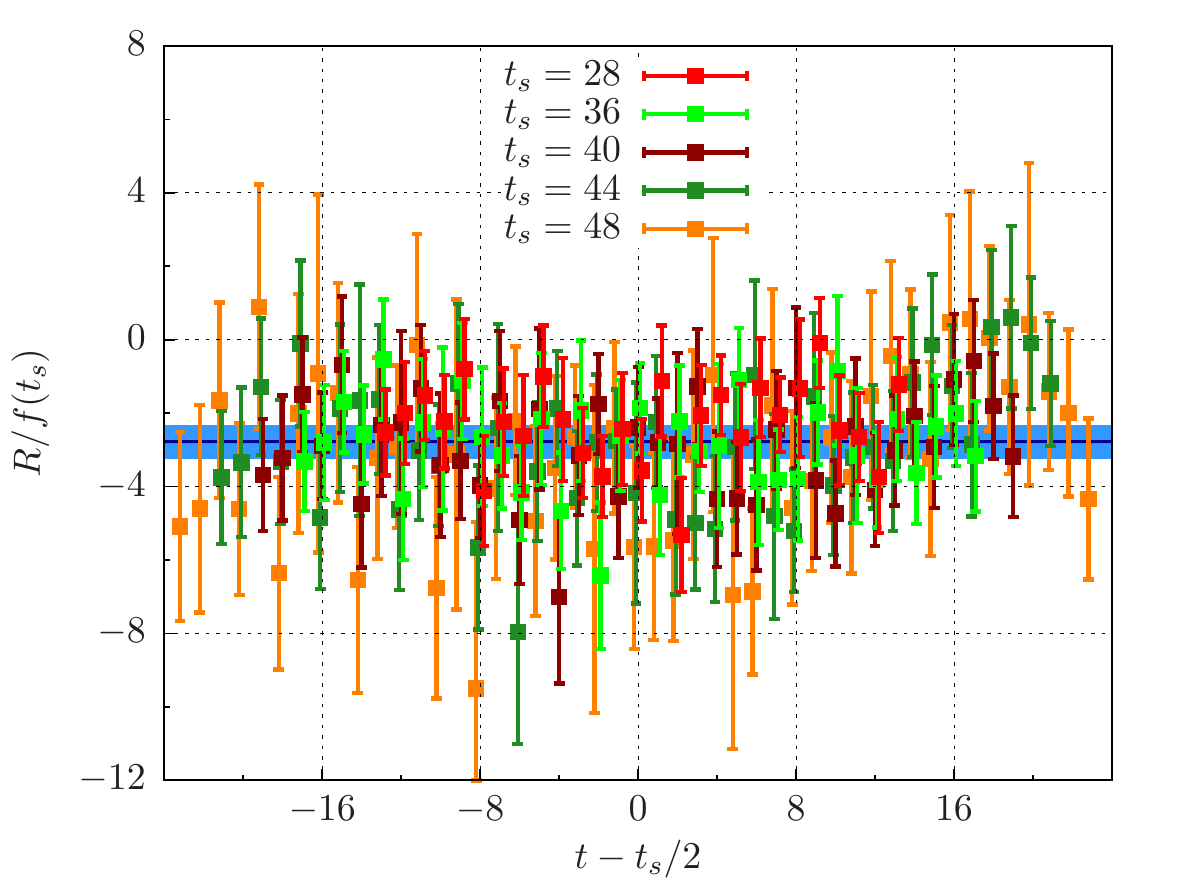}
\caption{Same as figure \ref{fig:mom0} shown for the F7
ensemble. The fit ranges in $t$ are listed in table
\ref{tab:fitrangest}.}
\label{fig:F7mom0}
\end{figure}

In figures \ref{fig:mom0} and \ref{fig:F7mom0} the ratios are 
plotted against the time $t$ of the
operator insertion at each value of the sink timeslice $t_s$ for E5 and for F7,
which has the lightest pion mass of all ensembles studied so far.
To account for the $t_s$-dependence (cf. equation
\eqref{eq:Rq20}) we have divided the ratios by the factor $f(t_s)$. Provided
that excited state contributions are sufficiently suppressed, one expects the
quantity $R(t,t_s,0,0)/f(t_s)$ to form plateaus in $t$ about $t_s/2$, which
are independent of $t_s$.
From the plots for the E5 ensemble one can easily see that the
ratios show a systematic trend as
the source-sink separation $t_s$ is increased, which is particularly apparent
in the case of the quark-connected contribution. At the same time one observes
that consistent plateaus are obtained when $t_s\geq24$. For the
quark-disconnected part the trend is somewhat obscured by the larger statistical
errors. The same $t_s$-behavior was already observed in the plateau values
shown in figure
\ref{fig:mom0ratiovsts}. The most likely explanation is the presence of excited
state contributions for $t_s<24$. 
In order to avoid a systematic bias, we have excluded ratios with $t_s<24$
from the subsequent analysis.
\begin{table}
      \begin{tabular}{|cr|l|}
	\hline
	\multicolumn{2}{|c|}{label} & $t_s$ values in global fit\\
	\hline\hline
	\multirow{2}{*}{E3 - E5} & connected & 	$24$, $28$, $32$\\
	   & disconnected & 	$24$, $26$, $28$, $30$, $32$\\
	\hline
	\multirow{2}{*}{F6, F7} & connected & 	$28$, $36$, $40$, $44$, $48$\\
	   & disconnected & 	$24$, $28$, $32$, $36$, $40$, $44$, $48$\\ 
	 \hline
      \end{tabular}
\caption{The values of $t_s$ that have been used in the global fits.}
\label{tab:fitranges}
\end{table}

The blue lines in the plots of figures \ref{fig:mom0} and
\ref{fig:F7mom0} show the results of
global fits to a constant within the plateau regions, applied to the data
computed for $t_s\geq24$. The values of $t_s$, that have been used for the
global fit are listed in table \ref{tab:fitranges}. Furthermore, in table
\ref{tab:fitrangest} we have compiled the fit ranges in $t$ applied to the
ensembles E5 and F7, which are shown in figures \ref{fig:mom0} and
\ref{fig:F7mom0}. The fit
result is proportional to the unnormalized scalar form factor
at vanishing momentum transfer,
\begin{equation}
\frac{R(t,t_s,0,0)}{f(t_s)} = 
\frac{\left<\pi(0)\right|\mathcal{O}_{\rm{S}}(0)
\left|\pi(0)\right>}{2m_\pi} = \frac{1}{2m_\pi} F_{_{\rm S}}^{\rm
bare}(Q^2=0)\,,
\end{equation}
where the pion mass $m_\pi=E_\pi(0)$ is known from the two-point function
$C_{2\textnormal{pt}}(t,0)$. 

We end this discussion with the observation that our method for the evaluation
of the quark-disconnected contribution can resolve the corresponding ratio
with good statistical accuracy at vanishing momentum transfer. The
plots on the right-hand side of figures \ref{fig:mom0} and
\ref{fig:F7mom0}, clearly show a good signal,
which differs from zero within several standard deviations.

\begin{figure}[h]
\centering
\includegraphics[scale=0.65]{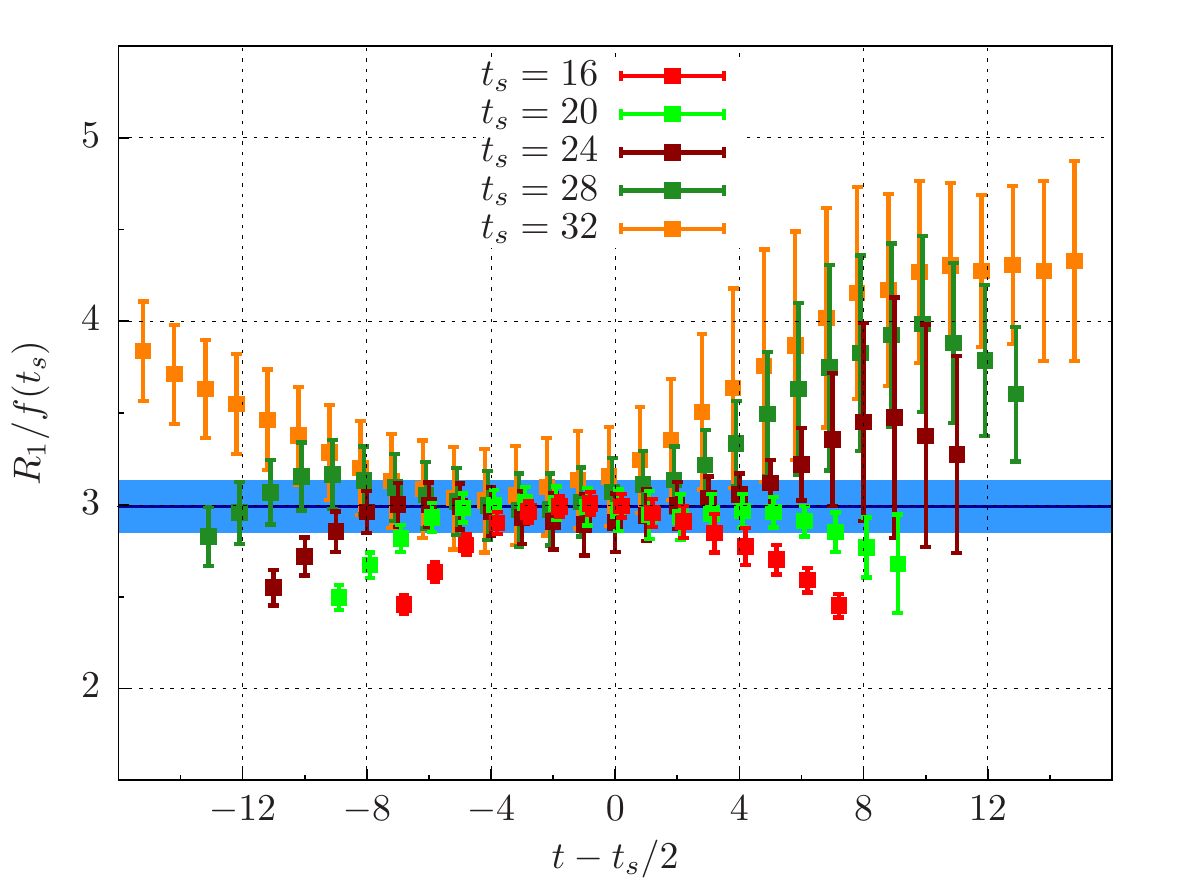} 
\includegraphics[scale=0.65]{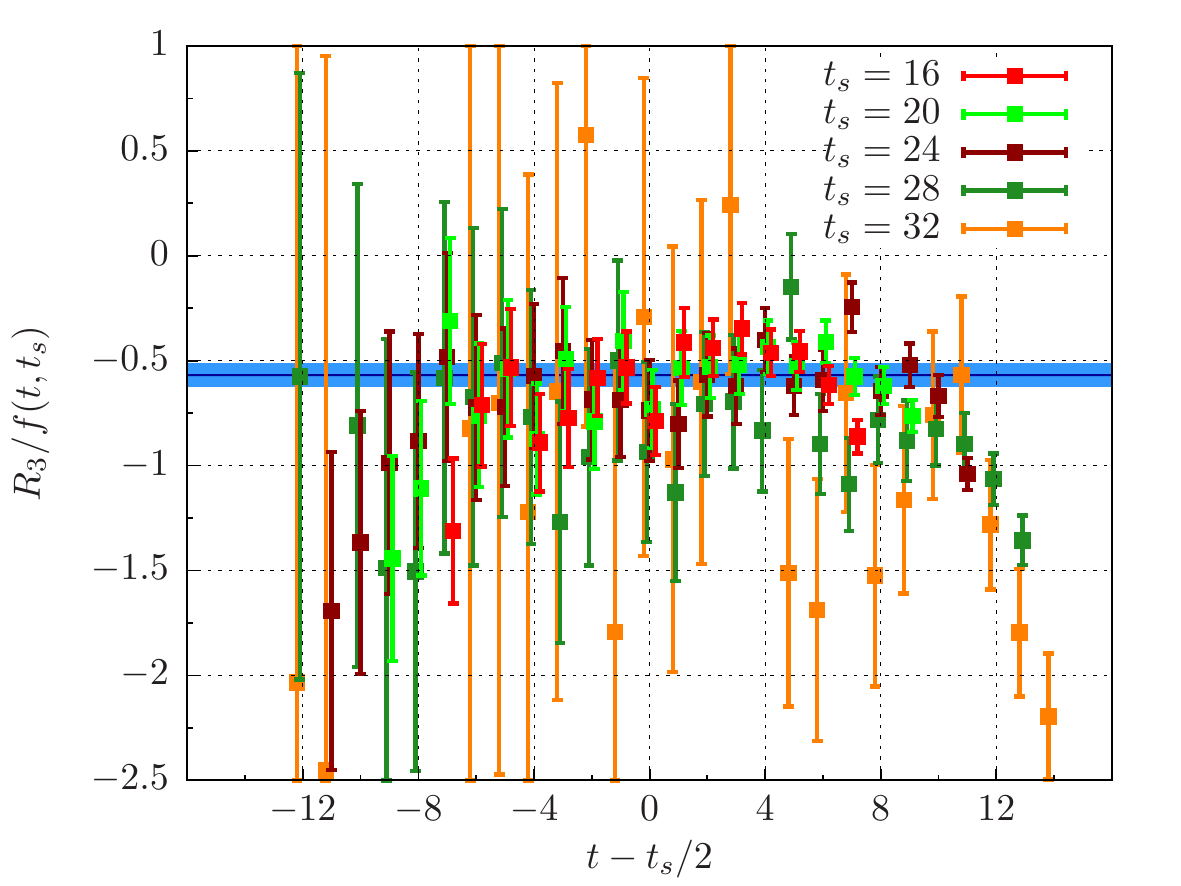}
\caption{Results for the ratios corrected by the time-dependence for
$Q^2=0.278$~GeV$^2$ for the E5 ensemble. The connected
contribution (smeared-local) is shown on the
left, the
disconnected (smeared-smeared) on the right. The blue lines indicate the results
of the global fit. The fit ranges in $t$ are listed in table
\ref{tab:fitrangest}.}
\label{fig:mom1}
\end{figure}
\begin{figure}[h]
\centering
\includegraphics[scale=0.65]{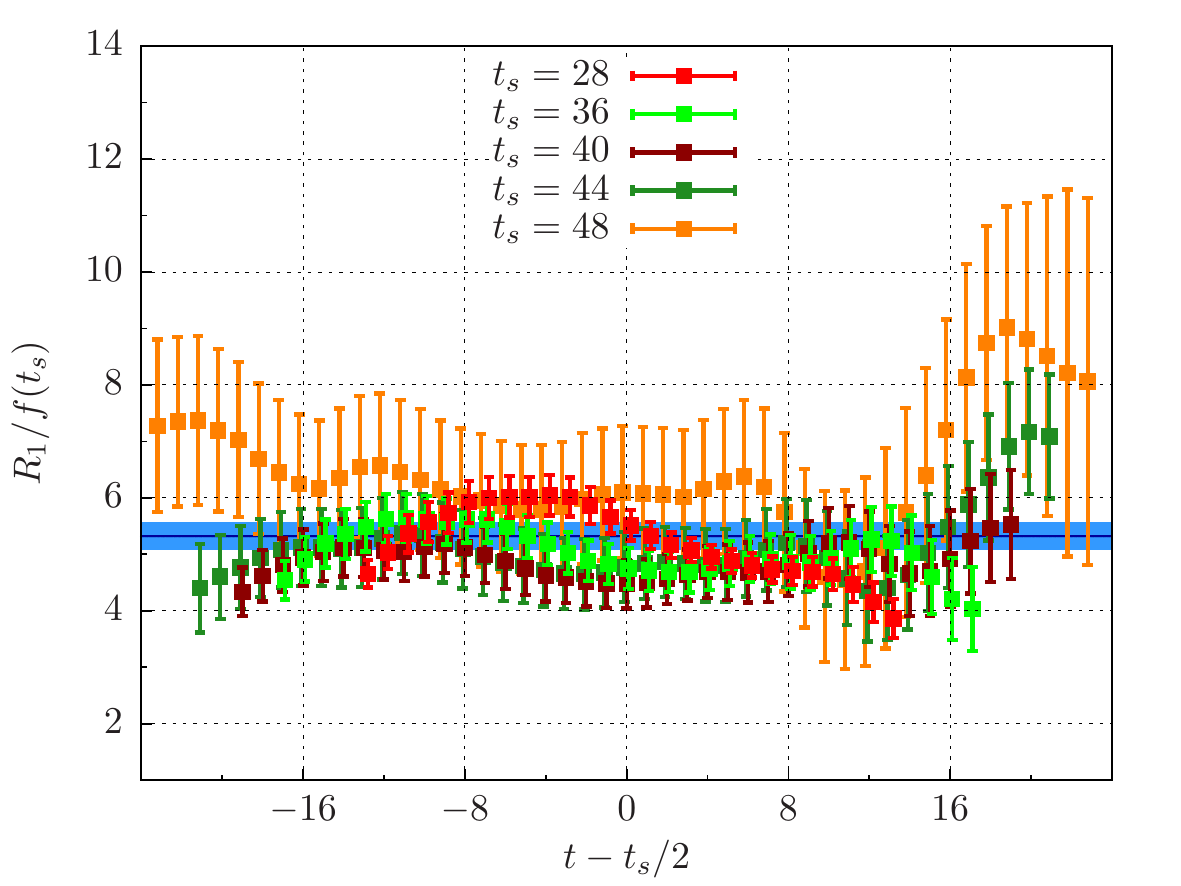} 
\includegraphics[scale=0.65]{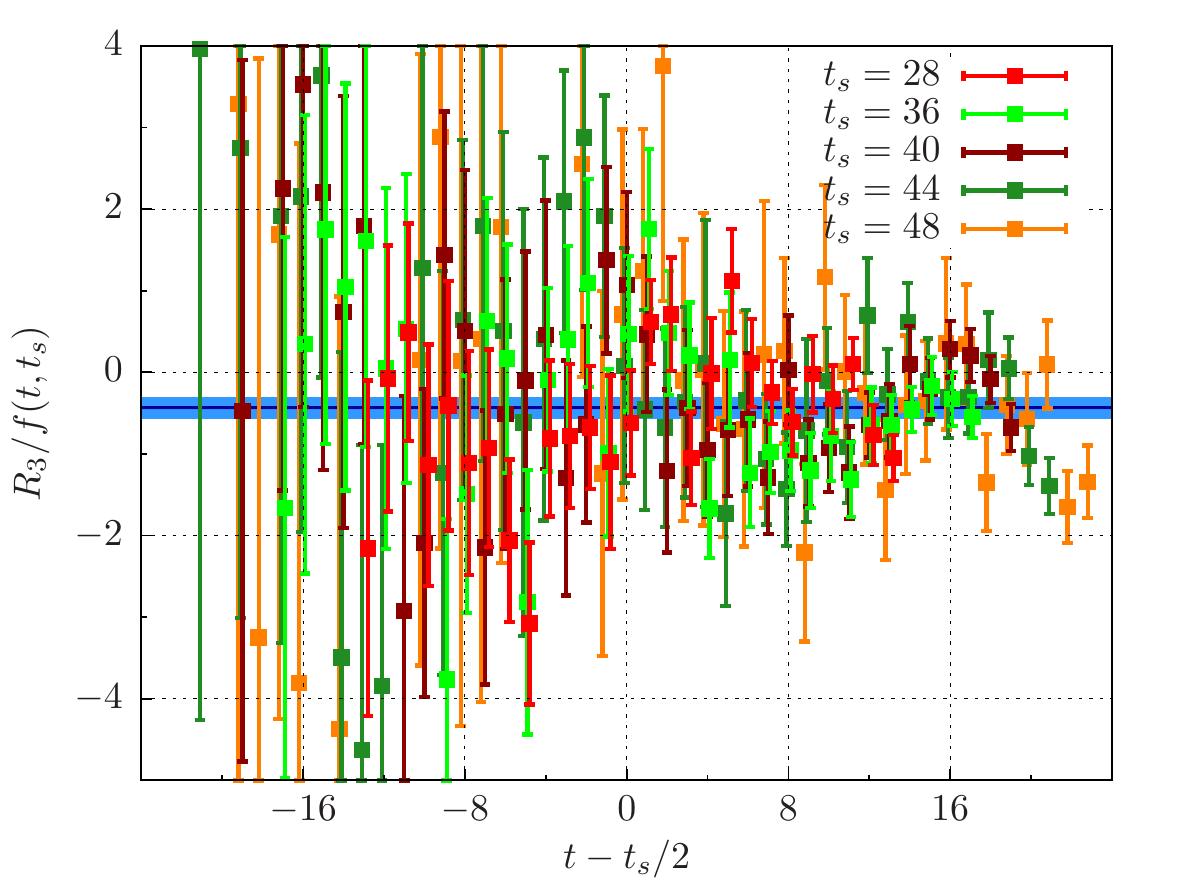}
\caption{Same as figure \ref{fig:mom1} shown for the F7
ensemble at $Q^2=0.121$~GeV$^2$. The fit ranges in $t$ are listed in table
\ref{tab:fitrangest}.}
\label{fig:F7mom1}
\end{figure}
\subsection{Ratios for $Q^2\neq0$}

As was mentioned above, we cannot employ twisted boundary conditions to study
the $Q^2$-dependence of the scalar form factor. Non-vanishing values of $Q^2$
are obtained by projecting the final-state pion and the insertion point of the
operator onto the values of $\vec{p}_f$ and $\vec{q}$, respectively.
On a finite lattice with spatial extent~$L$ the Fourier momenta are discrete,
and the smallest possible momentum is $\left|\vec{p}\right|=2\pi/L$. 
For the
ensembles E5 and F7, the minimum momentum transfer corresponds to
$Q^2=0.278$~GeV$^2$ and $Q^2=0.121$~GeV$^2$, respectively.

\begin{table}
     
\begin{tabular}{|c||p{0.3cm}cp{0.3cm}cp{0.3cm}|p{0.3cm}cp{0.3cm}cp{0.3cm}||p{
0.3cm}cp{0.3cm}cp{0.3cm}|p{0.3cm}cp{0.3cm}cp{0.3cm}|}
	\hline
	label &\multicolumn{5}{|c|}{connected $Q^2=0$}
&\multicolumn{5}{|c||}{disconnected $Q^2=0$}&\multicolumn{5}{|c|}{connected
$Q_1^2$}
&\multicolumn{5}{|c|}{disconnected $Q_1^2$} \\
&&$t_s$ && $t$&&&$t_s$ && $t$ &&&$t_s$ && $t$&&&$t_s$ && $t$ &\\ 
	\hline\hline
E5 && $24$ &&  $5-19$ &&& $24$ && $4-20$&
&& $24$ && $7-14$ &&& $24$ && $9-21$&\\
&& && &&& $26$ && $4-22$&
&& && &&& $26$ && $10-23$&\\
&& $28$ &&  $6-22$ &&& $28$ && $5-23$&
&& $28$ && $9-15$ &&& $28$ && $10-24$&\\
&& && &&& $30$ && $5-25$&
&& && &&& $30$ && $10-26$&\\
&& $32$ &&  $12-20$ &&& $32$ && $6-26$&
&& $32$ &&  $10-16$ &&& $32$ && $10-27$&\\
	\hline
F7 && && &&& $24$ && $4-20$&
&& && &&& $24$ && $3-21$&\\
&& $28$ &&  $6-15$ &&& $28$ && $5-23$&
&& $28$ &&  $6-12$ &&& $28$ && $4-25$&\\
&& && &&& $32$ && $6-26$&
&& && &&& $32$ && $7-29$&\\
&& $36$ &&  $13-28$ &&& $36$ && $7-29$&
&& $36$ &&  $6-13$ &&& $36$ && $12-32$&\\
&& $40$ &&  $18-25$ &&& $40$ && $8-32$&
&& $40$ &&  $18-27$ &&& $40$ && $16-35$&\\
&& $44$ &&  $18-28$ &&& $44$ && $8-36$&
&& $44$ &&  $18-27$ &&& $44$ && $20-39$&\\
&& $48$ &&  $19-29$ &&& $48$ && $8-40$&
&& $48$ &&  $18-28$ &&& $48$ && $23-43$&\\
	 \hline
      \end{tabular}
\caption{Values of the source-sink separation $t_s$ and the interval in $t$
used in the global fits to the connected and disconnected contributions to the
E5 and F7 ensembles.}
\label{tab:fitrangest}
\end{table}

To increase statistics for quark-disconnected contributions we have again used
four different source positions in the calculation of quark propagators. 
Additionally, we have averaged over all equivalent
momenta, e.g. $(0,0,2\pi/L)$, $(0,2\pi/L,0)$ and $(2\pi/L,0,0)$ for
the smallest non-zero value of $Q^2$.

As explained above, we use ratio $R_1$ of equation \eqref{eq:R1} for the
analysis for the
connected, and ratio $R_3$ of equation \eqref{eq:R3} for the disconnected
contribution.
Both ratios have known time-dependences which we can correct for. The ratios
with the time-dependence divided out are shown in figures
\ref{fig:mom1} and \ref{fig:F7mom1}, where
they are plotted against the operator insertion time $t$ for different values
of $t_s$. Within our statistical accuracy we do not see a trend in the data
computed for non-vanishing momentum transfer at different values of $t_s$,
unlike the case of $Q^2=0$ discussed earlier.  Nonetheless, we again exclude
the data with $t_s<24$ from the analysis, to be sure that systematic effects
from excited states are under control.
\par
As before, the blue lines in figures \ref{fig:mom1} and
\ref{fig:F7mom1} indicate the results from a
global fit to the plateau regions for different values of $t_s\geq24$. From
the fit results the scalar form factor for this momentum transfer can be
calculated,
\begin{align}
&R_1(t,t_s,\vec{p}_i,\vec{p}_f)/f(t_s) = 
\frac{\left<\pi(\vec{p}_i)\right|\mathcal{O}_{\rm{S}}(0)
\left|\pi(\vec{p}_f)\right>}{2\sqrt{E_\pi(\vec{p}_i)E_\pi(\vec{p}_f)}} =
\frac{1}{2\sqrt{E_\pi(\vec{p}_i)E_\pi(\vec{p}_f)}}\,F_{_{\rm S}}^{\rm
bare}(Q^2)\,, \\
&R_3(t,t_s,\vec{p}_i,\vec{p}_f)/f(t,t_s) = 
\frac{\left<\pi(\vec{p}_i)\right|\mathcal{O}_{\rm{S}}(0)
\left|\pi(\vec{p}_f)\right>}{2\sqrt{E_\pi(\vec{p}_i)E_\pi(\vec{p}_f)}} =
\frac{1}{2\sqrt{E_\pi(\vec{p}_i)E_\pi(\vec{p}_f)}}\,F_{_{\rm S}}^{\rm
bare}(Q^2)\,.
\end{align}
While the relative contribution of the quark-disconnected diagram to the form
factor is smaller compared to the case of vanishing momentum transfer,
$Q^2=0$, we note that our method is clearly able to resolve a signal.

In addition, we have included data for another momentum transfer, where the
final state
of the pion is projected to $\left|\vec{p}_f\right|=2\cdot2\pi/L$. The
corresponding pion two-point functions $C_{2\textrm{pt}}(t_s,\vec{p}_f)$, which
occur in the ratios, are fluctuating strongly, especially for larger values of
$t_s\lesssim T/2$, such that a reliable estimate for the form factor is not
possible using the two-point data themselves. Instead of dividing the
three-point function by $C_{2\textnormal{pt}}(t_s,\vec{p}_f)$ we use the fitted
two-point function in order to compute the ratios $R_1$ and $R_3$, which
reduces their statistical fluctuations.

\subsection{The $Q^2$ dependence of the form factor}
\begin{figure}[b]
\centering
\includegraphics[scale=0.65]{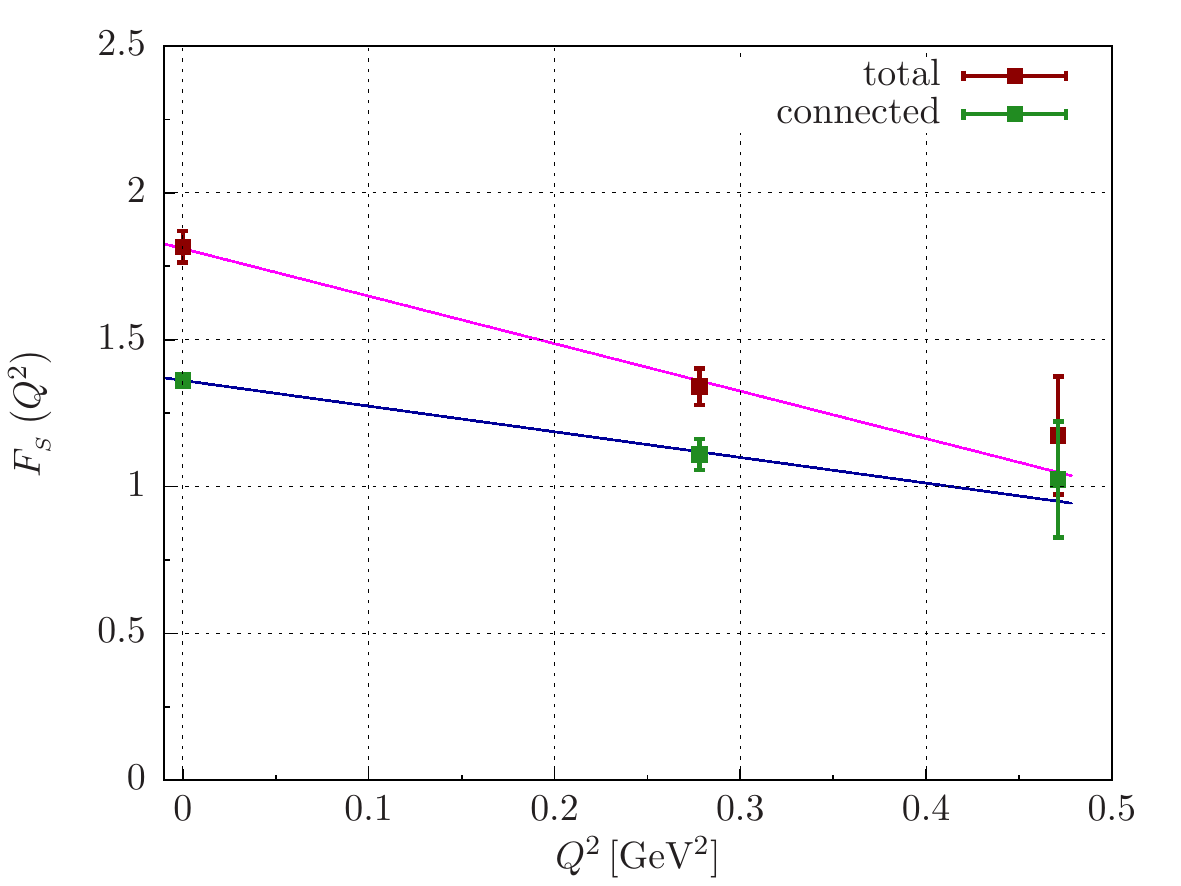}
\includegraphics[scale=0.65]{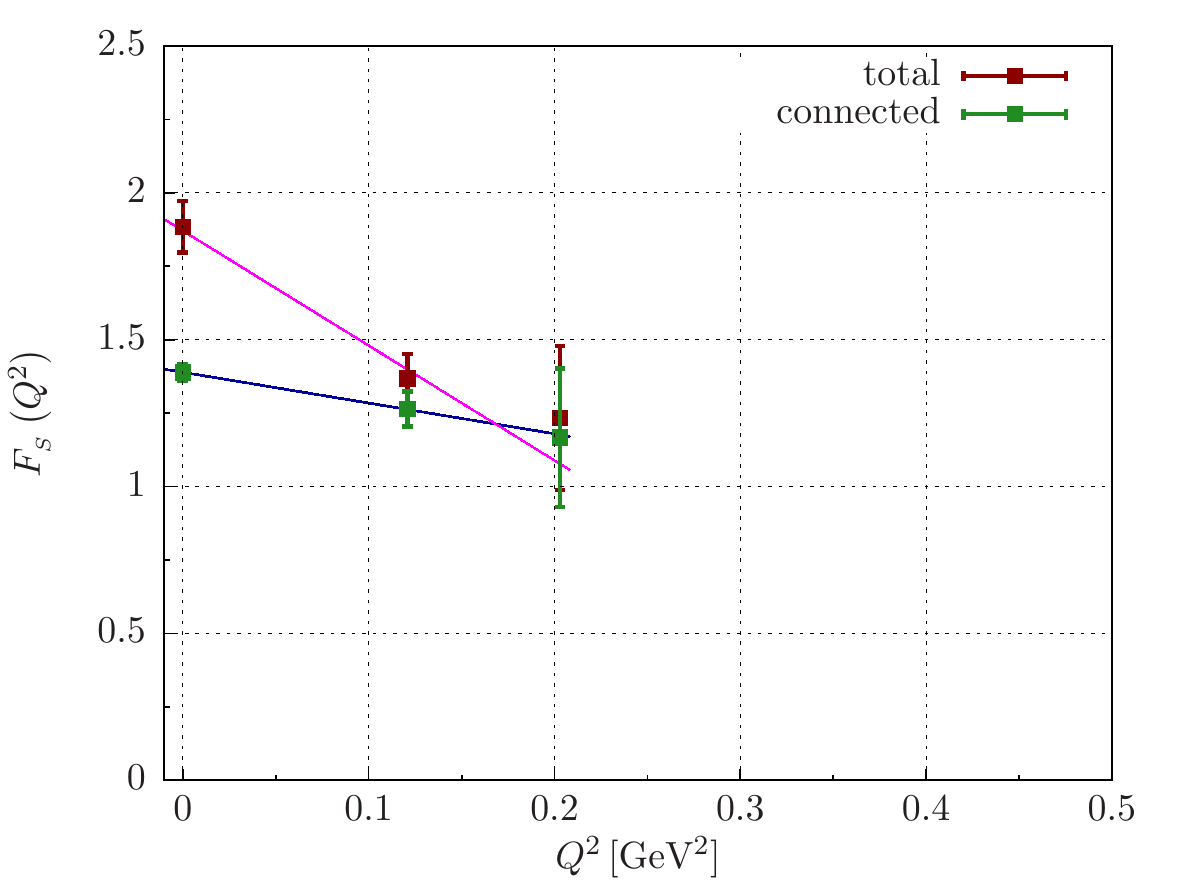}
\caption{The $Q^2$-dependence of the scalar form factor: on the left-hand side
E5 with a pion mass of $455$~MeV, on the right-hand side F7 with
$m_\pi=280$~MeV. The red points show the results for the total form factor
and the green points for the connected contribution only.}
\label{fig:q2}
\end{figure}

We briefly recall the definition of the scalar radius in terms of the scalar
form factor
\begin{equation}
 \left\langle r^2\right\rangle^\pi_{_{\rm S}} = -\frac{6}{F^\pi_{_{\rm S}}(0)}
                  \frac{\partial F^\pi_{_{\rm S}}(Q^2)}{\partial
Q^2}\Big|_{Q^2=0}\,.
\label{eq:scalarrdef}
\end{equation}
The scalar form factor admits an expansion, which has the general form
\begin{equation}
   F^\pi_{_{\rm S}}\left(Q^2\right) = F^\pi_{_{\rm S}}(0)
          \left(1 - \frac{1}{6}\left\langle r^2\right\rangle^\pi_{_{\rm S}} Q^2
                  + \mathcal{O}(Q^4) \right)\,,
\label{eq:Q2dependence}
\end{equation}
and which is consistent with the definition \eqref{eq:scalarrdef} of the scalar
radius. 

In practice, the slope at $Q^2=0$ is difficult to determine on the lattice in
a model-independent way. Usually one fits the lattice data for form factors
obtained at a few discrete values of $Q^2$ to some phenomenological model such
as vector meson dominance. In the case of the pion vector form factor, which
is amenable to the use of twisted boundary conditions, it is possible to tune
$Q^2$ so as to generate a high density of data points in the immediate
vicinity of $Q^2=0$ from which the slope can be extracted without any model
assumptions \cite{Brandt:2013dua}.

Here we must resort to a more naive treatment, since twisted boundary
conditions cannot be used to evaluate the quark-disconnected contribution, so
that the resolution in $Q^2$ is only quite rough. As a consequence, we
estimate the scalar radius from a linear fit over a relatively broad
interval in $Q^2$, using three data points only. However, we compare different
fit ans\"atze in an attempt to investigate the systematics of this procedure.

In figure \ref{fig:q2} we show the  $Q^2$-dependence for the ensembles E5 and
F7. Both plots show the total form factor and the results
obtained when the
disconnected contributions are neglected. According to
\eqref{eq:Q2dependence}
a linear function was fitted to the data to estimate the scalar
radius. For both ensembles shown here the descending slope of the linear curves
is clearly steeper for the total form factor than for the connected part
only. This stresses the importance of including the disconnected diagram for
determining the scalar radius.

\begin{figure}[t]
\centering
\includegraphics[scale=0.75]{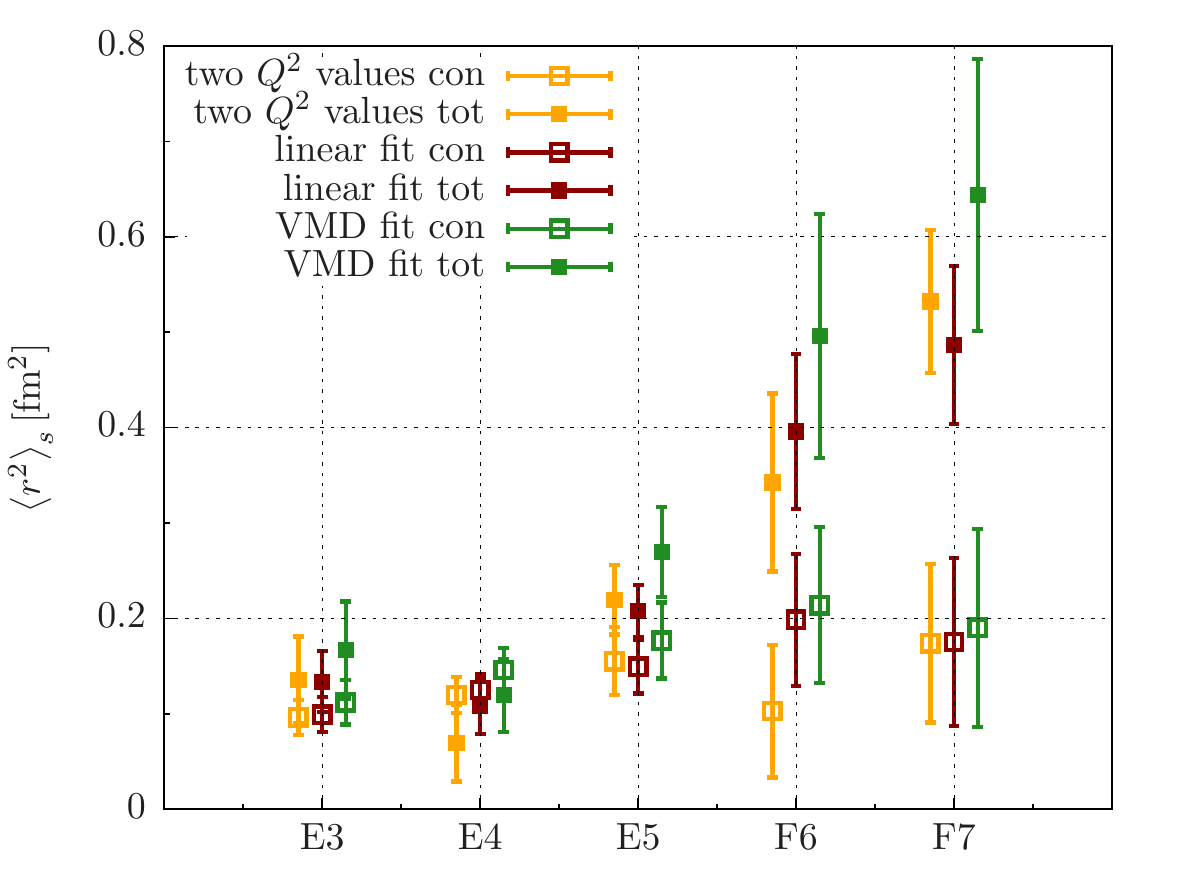}
\caption{A comparison of different descriptions of the form factor
data: a linear interpolation (yellow) at the two smallest values of
$Q^2$, and both a linear (red) and a VMD-inspired (green) fit to the
three smallest values of $Q^2$, with the radii resulting from
considering only the connected part and the complete three-point
function shown using open and filled symbols, respectively.}
\label{fig:compare}
\end{figure}
%
%
%

%
%
%
%

\begin{table}[b]
 \centering

      \begin{tabular}{|cr||c|c|c|c|c||c|}
\hline
&&	$F^\pi_{_{\rm S}}\left(0\right)$ & $Q_1^2
\left[\textnormal{GeV}^2\right]$ & $F^\pi_{_{\rm
S}}\left(Q_1^2\right)$ & $Q_2^2
\left[\textnormal{GeV}^2\right]$ &
$F^\pi_{_{\rm S}}\left(Q_2^2\right)$ & $\left\langle
r^2\right\rangle^\pi_{_{\rm S}} \left[\textnormal{fm}^2\right]$ \\
\hline\hline
\multirow{2}{*}{E3} & connected & $1.39\pm 0.01$ &\multirow{2}{*}{$0.319$}&
$1.20\pm0.03$ &\multirow{2}{*}{$0.565$}& $1.02\pm0.12$ &
$0.099\pm0.018$\\
		  & total & $1.97\pm0.11$ && $1.61\pm0.07$ && $1.33\pm0.14$ &
$0.134\pm0.032$\\
\hline
\multirow{2}{*}{E4} & connected & $1.39\pm0.01$ &\multirow{2}{*}{$0.311$}&
$1.17\pm0.04$  &\multirow{2}{*}{$0.548$}& $0.93\pm0.11$
& $0.125\pm0.017$\\
		  & total & $1.88\pm0.09$ && $1.70\pm0.06$ && $1.38\pm0.13$ &
$0.208\pm0.027$\\
\hline
\multirow{2}{*}{E5} & connected & $1.36\pm0.01$ &\multirow{2}{*}{$0.278$}&
$1.11\pm0.05$ &\multirow{2}{*}{$0.471$}& $1.02\pm0.19$
& $0.149\pm0.028$\\
		  & total & $1.82\pm0.05$ && $1.34\pm0.06$ && $1.17\pm0.20$ &
$0.208\pm0.027$\\
\hline
\multirow{2}{*}{F6} & connected & $1.44\pm0.03$ &\multirow{2}{*}{$0.128$}&
$1.36\pm0.07$ &\multirow{2}{*}{$0.221$}& $1.02\pm0.14$
& $0.197\pm0.069$\\
		  & total & $1.97\pm0.10$ && $1.60\pm0.08$ && $1.17\pm0.16$ &
$0.396\pm0.081$\\
\hline
\multirow{2}{*}{F7} & connected & $1.39\pm0.03$ &\multirow{2}{*}{$0.121$}&
$1.26\pm0.06$ &\multirow{2}{*}{$0.203$}& $1.17\pm0.23$
& $0.175\pm0.088$\\
		  & total & $1.88\pm0.09$ && $1.37\pm0.08$ && $1.23\pm0.24$
& $0.487\pm0.083$\\
\hline
      \end{tabular}
\caption{Numerical results of the scalar pion form factor $F^\pi_{_{\rm
S}}\left(Q^2\right)$ for three different momentum transfers $Q^2$ and the
results for the scalar radius $\left\langle
r^2\right\rangle^\pi_{_{\rm S}}$ as determined from an uncorrelated linear fit.}
\label{tab:results}
\end{table}

\normalsize

For all ensembles studied so far, we find the results for the three different
$Q^2$ to be consistent with a linear $Q^2$ dependence within their statistical
errors. In order to investigate the systematic effect in the determination of
the scalar radius arising from the ansatz for the $Q^2$ dependence, we have
compared the linear fit to a VMD-inspired fit of the form $1/(1 + Q^2/M^2)^2$
as well as a linear interpolation using only the two smallest $Q^2$ values. As
can be inferred from figure \ref{fig:compare} no statistically significant
effect in the determination of $\left\langle r^2\right\rangle^\pi_{_{\rm S}}$
arising from the use of different ans\"atze is observed. This indicates that
any possible curvature contained in the data cannot be resolved at the current
level of statistical accuracy.\par
We choose the linear fit as a reasonable
compromise between achieving a well-motivated description of the data and
keeping the statistical error of the fitted radius in check. The results for
the form factor and the scalar radius from the linear fit are summarized in
table \ref{tab:results}.

\subsection{Chiral extrapolation}
\label{subsec:extra}

Since our simulations of the scalar radius have been performed with pion masses
larger than the physical mass $m_\pi>m_{\pi,\textnormal{phys}}$, we have
to perform a chiral extrapolation. In chiral perturbation theory at NLO the
scalar radius of the pion is
\cite{Gasser:1983yg,Gasser:1990bv,Bijnens:1998fm}
\begin{equation}
\left\langle r^2\right\rangle^\pi_{_{\rm S}} = \frac{1}{(4\pi F)^2}
\left(-\frac{13}{2}\right)
+ \frac{6}{(4\pi F)^2}\left[\overline{\ell}_4
+ \ln\left(\frac{m_{\pi,phys}^2}{m_\pi^2}\right)\right]
\label{eq:NLO}
\end{equation}
where $F=92.2$~MeV
\cite{PDG:2012}
is the pion decay constant.
\begin{figure}[b]
 \begin{center}
  \includegraphics[scale=0.75]{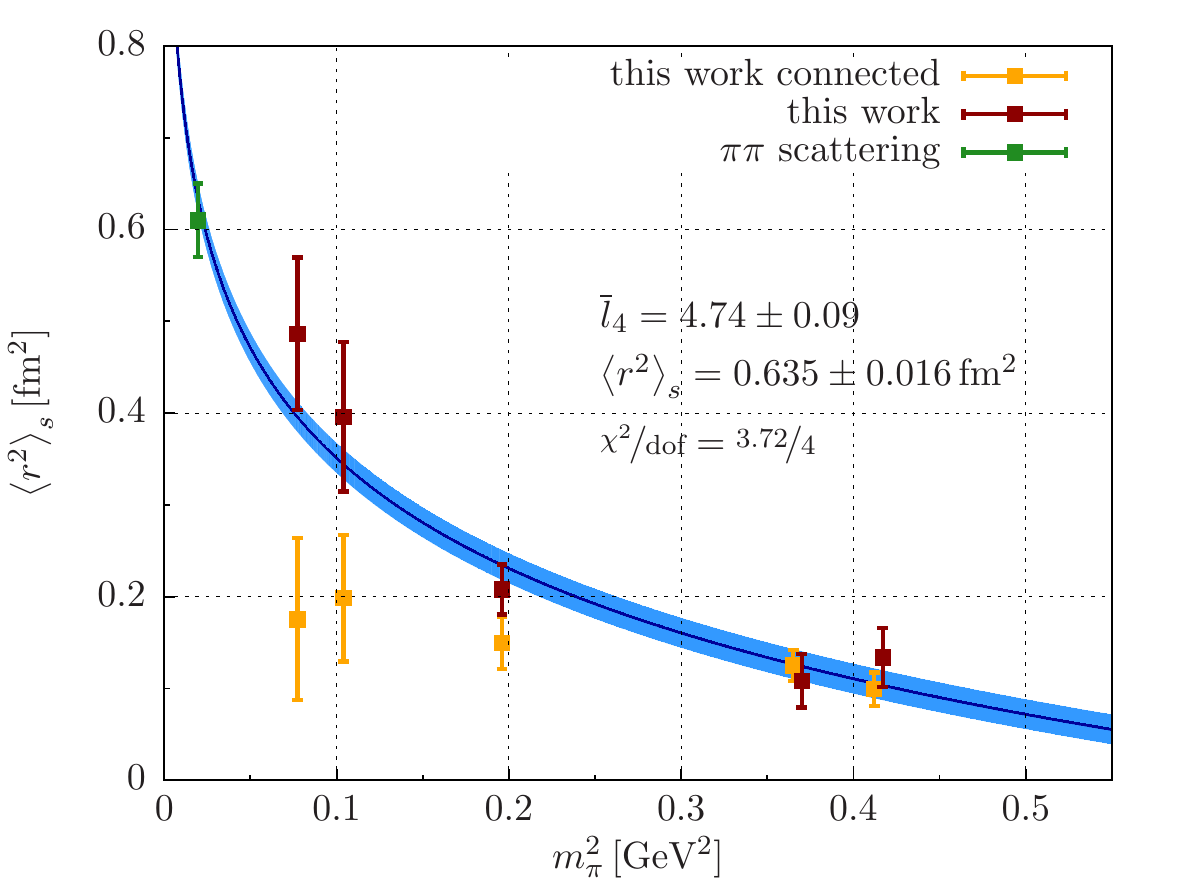}
 \end{center}
\caption{The $m_\pi^2$-dependence of the scalar radius. The blue band is a
    fit to the lattice data obtained from both quark-connected and
    -disconnected diagrams.}
\label{fig:rsqvsmpisq}
\end{figure}

In figure \ref{fig:rsqvsmpisq} the values obtained for
$\left\langle r^2\right\rangle^\pi_{_{\rm S}}$ are plotted against the square
of the pion mass, $m_\pi^2$. The point shown at the physical pion mass is the
value obtained from $\pi\pi$-scattering
\cite{Colangelo:2001df}.
The expression \eqref{eq:NLO} from NLO \xPT\ has been fitted to the data
and the obtained curve is shown in blue. This fit allows a determination of
the low energy constant $\overline{\ell}_4$ for which we find
 $\overline{\ell}_4=4.74\pm0.09$, where the error is only
statistical.

This is in excellent agreement with the result of ref.
\cite{Brandt:2013dua},
which was extracted from chiral fits to the pseudoscalar decay constant
computed on the CLS ensembles at three different lattice spacings. The result
for the scalar radius at physical pion mass obtained from our NLO fit is
\begin{equation}
 \left\langle r^2\right\rangle^\pi_{_{\rm S}} = 0.635\pm0.016\
\textnormal{fm}^2\,,
\end{equation}
which agrees very well with the $\pi\pi$-scattering value
$\left\langle r^2\right\rangle^\pi_{_{\rm S}}=0.61\pm0.04$~fm$^2$ reported in
ref.
\cite{Colangelo:2001df}.
In figure \ref{fig:rsqvsmpisq} one can see that our data are well described by
\xPT\ at NLO. As already indicated in figure \ref{fig:q2}, the
quark-disconnected contribution to the scalar radius of the pion is not
negligible. The yellow points in figure \ref{fig:rsqvsmpisq} show the data
obtained from the connected contribution only. For the ensembles analyzed so
far, we find that the disconnected contribution to the scalar radius becomes
more important as the pion mass approaches its physical value. Clearly,
neglecting the disconnected diagram fails to reproduce the phenomenological
expectation for the scalar radius.

These findings differ from the results obtained by the JLQCD and
TWQCD collaborations
\cite{Aoki:2009qn},
where no significant pion mass dependence of the scalar radius was observed.
The reason for this discrepancy is presently unknown. Here we only comment
that the two simulations in question differ substantially regarding the value
of the lattice spacing, the minimum value of $m_\pi L$, and the type of
fermionic discretization. It should also be noted that the contribution of
quark-disconnected diagrams in
\cite{Aoki:2009qn},
though significant, was observed to be much smaller than in our study.
Clearly, more work is needed to investigate the systematics of these
calculations. To this end we will add more ensembles at smaller pion masses
and different lattice spacings.

\section{Conclusions}
\label{sec:conclusions}

The combination of the hopping parameter expansion with the use of stochastic
sources provides a powerful means for estimating quark-disconnected
contributions to hadronic form factors. We have been able to obtain a clearly
non-vanishing signal for the scalar form factor of the pion both at $Q^2=0$
(where there is a large subtraction of the vacuum contribution) and
at non-vanishing momentum transfer, where the
correlation functions become intrinsically noisy.

We find that the disconnected contribution to the scalar form factor is
not negligible, and that indeed the purely connected part of the form factor
fails to reproduce the expected logarithmic behaviour of the pion scalar radius
as a function of the pion mass. This is in qualitative agreement with what
has been found in partially quenched \xPT\
\cite{Juttner:2011ur}.
From our determination of the pion scalar radius, we can derive a lattice
estimate of the low-energy constant $\overline{\ell}_4=4.74\pm0.09$, which is in
fair
agreement with the phenomenological estimate
\cite{Colangelo:2001df}
$\bar{\ell}_4=4.4\pm 0.2$ based on the analysis of $\pi\pi$-scattering
amplitudes.

The present study is based on a single, albeit rather fine, lattice spacing.
It is therefore important to repeat this study on ensembles with different
values of the lattice spacing, to estimate the size of discretization effects
and perform an extrapolation to the continuum limit. Another potential source
of systematic errors are finite-volume effects. While all of our lattices
satisfy $M_\pi L\ge 4$, it is desirable to include further, even larger,
lattice volumes to ensure that finite-volume effects are indeed fully under
control.

Another source of systematic error in the determination of the pion scalar
radius, and hence of $\bar{\ell}_4$, is the simple linear fit
used to estimate the derivative of the scalar form factor at vanishing $Q^2$.
It would be highly desirable to augment this somewhat naive approximation
by using partially twisted boundary conditions for the connected part along
the lines of
\cite{Jiang:2006gna,Brandt:2013mb}.
Unfortunately this method is fundamentally inapplicable to  the disconnected
part, where the same quark propagator connects to the operator insertion on
both sides, and some interpolation will necessarily be required in this case.
However, all our data are consistent with a linear $Q^2$ dependence,
and any possible curvature cannot be resolved with our current accuracy.

Finally, another potential for systematic error lies in the use of NLO \xPT\
formulae, which may not always give a good description of pion form factors
\cite{Brandt:2013dua}.
The ability of the NLO expressions to describe the numerical data crucially
depends on the overall accuracy of the latter. If the the statistical errors in
the determinations of the scalar form factor and radius can be substantially
decreased, one may have to resort to \xPT\ at NNLO.

\begin{acknowledgments}
We acknowledge useful discussions with Andreas J\"uttner, Bastian Brandt and
Harvey B.~Meyer.  Our calculations were performed on the ``Wilson'' HPC
Cluster at the Institute for Nuclear Physics, University of Mainz. We thank
Dalibor Djukanovic and Christian Seiwerth for technical support. We are
grateful for computer time allocated to project HMZ21 on the BlueGene
computers ``JUGENE'' and ``JUQUEEN'' at NIC, J\"ulich. This research has been
supported in part by the DFG in the SFB~1044. We are grateful to our
colleagues in the CLS initiative for sharing ensembles.
\end{acknowledgments}
\bibliographystyle{h-physrev4}
\bibliography{scalarformfactor}

\begin{thebibliography}{10}

\bibitem{Gasser:1983yg}
J.~Gasser and H.~Leutwyler,
\newblock Annals Phys. {\bf 158}, 142 (1984).

\bibitem{Gasser:1984gg}
J.~Gasser and H.~Leutwyler,
\newblock Nucl. Phys. {\bf B250}, 465 (1985).

\bibitem{Durr:2010aw}
S.~D{\"u}rr {\em et~al.},
\newblock JHEP {\bf 1108}, 148 (2011), [1011.2711].

\bibitem{Durr:2010vn}
S.~D{\"u}rr {\em et~al.},
\newblock Phys. Lett. {\bf B701}, 265 (2011), [1011.2403].

\bibitem{Aoki:2009ix}
PACS-CS Collaboration, S.~Aoki {\em et~al.},
\newblock Phys. Rev. {\bf D81}, 074503 (2010), [0911.2561].

\bibitem{Heitger:2000ay}
ALPHA Collaboration, J.~Heitger, R.~Sommer and H.~Wittig,
\newblock Nucl. Phys. {\bf B588}, 377 (2000), [hep-lat/0006026].

\bibitem{Giusti:2003iq}
L.~Giusti, P.~Hernandez, M.~Laine, P.~Weisz and H.~Wittig,
\newblock JHEP {\bf 0401}, 003 (2004), [hep-lat/0312012].

\bibitem{Giusti:2004yp}
L.~Giusti, P.~Hernandez, M.~Laine, P.~Weisz and H.~Wittig,
\newblock JHEP {\bf 0404}, 013 (2004), [hep-lat/0402002].

\bibitem{Gattringer:2005ij}
Bern-Graz-Regensburg (BGR) Collaboration, C.~Gattringer, P.~Huber and C.~Lang,
\newblock Phys. Rev. {\bf D72}, 094510 (2005), [hep-lat/0509003].

\bibitem{Hasenfratz:2008ce}
A.~Hasenfratz, R.~Hoffmann and S.~Schaefer,
\newblock Phys. Rev. {\bf D78}, 054511 (2008), [0806.4586].

\bibitem{Beane:2011zm}
S.~Beane {\em et~al.},
\newblock Phys. Rev. {\bf D86}, 094509 (2012), [1108.1380].

\bibitem{Bernardoni:2011fx}
F.~Bernardoni, J.~Bulava and R.~Sommer,
\newblock PoS {\bf LATTICE2011}, 095 (2011), [1111.4351].

\bibitem{Damgaard:2012gy}
P.~Damgaard, U.~Heller and K.~Splittorff,
\newblock Phys. Rev. {\bf D86}, 094502 (2012), [1206.4786].

\bibitem{Borsanyi:2012zv}
S.~Borsanyi {\em et~al.},
\newblock Phys.Rev. {\bf D88}, 014513 (2013), [1205.0788].

\bibitem{Herdoiza:2013sla}
G.~Herdoiza, K.~Jansen, C.~Michael, K.~Ottnad and C.~Urbach,
\newblock JHEP {\bf 1305}, 038 (2013), [1303.3516].

\bibitem{Capitani:2005ce}
Bern-Graz-Regensburg (BGR) Collaboration, S.~Capitani, C.~Gattringer and
  C.~Lang,
\newblock Phys. Rev. {\bf D73}, 034505 (2006), [hep-lat/0511040].

\bibitem{Brommel:2006ww}
QCDSF/UKQCD Collaboration, D.~Br{\"o}mmel {\em et~al.},
\newblock Eur. Phys. J. {\bf C51}, 335 (2007), [hep-lat/0608021].

\bibitem{Jiang:2006gna}
F.-J. Jiang and B.~Tiburzi,
\newblock Phys. Lett. {\bf B645}, 314 (2007), [hep-lat/0610103].

\bibitem{Kaneko:2007nf}
JLQCD Collaboration, T.~Kaneko {\em et~al.},
\newblock PoS {\bf LAT2007}, 148 (2007), [0710.2390].

\bibitem{Alexandrou:2007pn}
C.~Alexandrou and G.~Koutsou,
\newblock PoS {\bf LAT2007}, 150 (2007), [0710.2441].

\bibitem{Boyle:2008yd}
P.~Boyle {\em et~al.},
\newblock JHEP {\bf 0807}, 112 (2008), [0804.3971].

\bibitem{Aoki:2009qn}
JLQCD and TWQCD Collaborations, S.~Aoki {\em et~al.},
\newblock Phys. Rev. {\bf D80}, 034508 (2009), [0905.2465].

\bibitem{Nguyen:2011ek}
O.~H. Nguyen, K.-I. Ishikawa, A.~Ukawa and N.~Ukita,
\newblock JHEP {\bf 1104}, 122 (2011), [1102.3652].

\bibitem{Fukaya:2012dla}
JLQCD Collaboration, H.~Fukaya {\em et~al.},
\newblock PoS {\bf LATTICE2012}, 198 (2012), [1211.0743].

\bibitem{Brandt:2013mb}
B.~B. Brandt, A.~J{\"u}ttner and H.~Wittig,
\newblock PoS {\bf ConfinementX}, 112 (2012), [1301.3513].

\bibitem{Gasser:1983kx}
J.~Gasser and H.~Leutwyler,
\newblock Phys. Lett. {\bf B125}, 325 (1983).

\bibitem{Donoghue:1990xh}
J.~F. Donoghue, J.~Gasser and H.~Leutwyler,
\newblock Nucl. Phys. {\bf B343}, 341 (1990).

\bibitem{Gasser:1990bv}
J.~Gasser and U.~G. Meissner,
\newblock Nucl. Phys. {\bf B357}, 90 (1991).

\bibitem{Moussallam:1999aq}
B.~Moussallam,
\newblock Eur. Phys. J. {\bf C14}, 111 (2000), [hep-ph/9909292].

\bibitem{Colangelo:2001df}
G.~Colangelo, J.~Gasser and H.~Leutwyler,
\newblock Nucl. Phys. {\bf B603}, 125 (2001), [hep-ph/0103088].

\bibitem{Juttner:2011ur}
A.~J{\"u}ttner,
\newblock JHEP {\bf 1201}, 007 (2012), [1110.4859].

\bibitem{Juttner:2012xs}
A.~J{\"u}ttner,
\newblock PoS {\bf LATTICE2012}, 196 (2012), [1212.2559].

\bibitem{Bitar:1988bb}
K.~Bitar, A.~Kennedy, R.~Horsley, S.~Meyer and P.~Rossi,
\newblock Nucl. Phys. {\bf B313}, 348 (1989).

\bibitem{Neff:2001zr}
H.~Neff, N.~Eicker, T.~Lippert, J.~W. Negele and K.~Schilling,
\newblock Phys. Rev. {\bf D64}, 114509 (2001), [hep-lat/0106016].

\bibitem{Bali:2005fu}
SESAM Collaboration, G.~S. Bali, H.~Neff, T.~Duessel, T.~Lippert and
  K.~Schilling,
\newblock Phys. Rev. {\bf D71}, 114513 (2005), [hep-lat/0505012].

\bibitem{Thron:1997iy}
C.~Thron, S.~Dong, K.~Liu and H.~Ying,
\newblock Phys. Rev. {\bf D57}, 1642 (1998), [hep-lat/9707001].

\bibitem{Collins:2007mh}
S.~Collins, G.~Bali and A.~Sch{\"a}fer,
\newblock PoS {\bf LAT2007}, 141 (2007), [0709.3217].

\bibitem{Gulpers:2012kd}
V.~G{\"u}lpers, G.~von Hippel and H.~Wittig,
\newblock PoS {\bf LATTICE2012}, 181 (2012).

\bibitem{Sheikholeslami:1985ij}
B.~Sheikholeslami and R.~Wohlert,
\newblock Nucl. Phys. {\bf B259}, 572 (1985).

\bibitem{Luscher:1996sc}
M.~L{\"u}scher, S.~Sint, R.~Sommer and P.~Weisz,
\newblock Nucl. Phys. {\bf B478}, 365 (1996), [hep-lat/9605038].

\bibitem{Luscher:2005rx}
M.~L{\"u}scher,
\newblock Comput.Phys.Commun. {\bf 165}, 199 (2005), [hep-lat/0409106].

\bibitem{Luscher:2007es}
M.~L{\"u}scher,
\newblock JHEP {\bf 0712}, 011 (2007), [0710.5417].

\bibitem{Jansen:1998mx}
ALPHA Collaboration, K.~Jansen and R.~Sommer,
\newblock Nucl. Phys. {\bf B530}, 185 (1998), [hep-lat/9803017].

\bibitem{Capitani:2011fg}
S.~Capitani, M.~Della~Morte, G.~von Hippel, B.~Knippschild and H.~Wittig,
\newblock PoS {\bf LATTICE2011}, 145 (2011), [1110.6365].

\bibitem{Fritzsch:2012wq}
P.~Fritzsch {\em et~al.},
\newblock Nucl. Phys. {\bf B865}, 397 (2012), [1205.5380].

\bibitem{Martinelli:1988rr}
G.~Martinelli and C.~T. Sachrajda,
\newblock Nucl. Phys. {\bf B316}, 355 (1989).

\bibitem{Bali:2009hu}
G.~S. Bali, S.~Collins and A.~Sch{\"a}fer,
\newblock Comput.Phys.Commun. {\bf 181}, 1570 (2010), [0910.3970].

\bibitem{Gulpers:Diplom}
V.~G{\"u}lpers,
\newblock Diploma thesis, JGU Mainz, 2011,
\newblock {URL}:
  \url{http://wwwkph.kph.uni-mainz.de/T//pub/diploma/Dipl_Th_Guelpers.pdf}.

\bibitem{Boyle:2007wg}
P.~Boyle, J.~Flynn, A.~J{\"u}ttner, C.~Sachrajda and J.~Zanotti,
\newblock JHEP {\bf 0705}, 016 (2007), [hep-lat/0703005].

\bibitem{Bonnet:2004fr}
LHP Collaboration, F.~D. Bonnet, R.~G. Edwards, G.~T. Fleming, R.~Lewis and
  D.~G. Richards,
\newblock Phys. Rev. {\bf D72}, 054506 (2005), [hep-lat/0411028].

\bibitem{Gusken:1989ad}
S.~G{\"u}sken {\em et~al.},
\newblock Phys. Lett. {\bf B227}, 266 (1989).

\bibitem{Alexandrou:1990dq}
C.~Alexandrou, F.~Jegerlehner, S.~G{\"u}sken, K.~Schilling and R.~Sommer,
\newblock Phys. Lett. {\bf B256}, 60 (1991).

\bibitem{Allton:1993wc}
UKQCD Collaboration, C.~Allton {\em et~al.},
\newblock Phys. Rev. {\bf D47}, 5128 (1993), [hep-lat/9303009].

\bibitem{Bedaque:2004kc}
P.~F. Bedaque,
\newblock Phys.Lett. {\bf B593}, 82 (2004), [nucl-th/0402051].

\bibitem{Sachrajda:2004mi}
C.~Sachrajda and G.~Villadoro,
\newblock Phys.Lett. {\bf B609}, 73 (2005), [hep-lat/0411033].

\bibitem{Flynn:2005in}
UKQCD, J.~Flynn, A.~J{\"u}ttner and C.~Sachrajda,
\newblock Phys.Lett. {\bf B632}, 313 (2006), [hep-lat/0506016].

\bibitem{deDivitiis:2004kq}
G.~de~Divitiis, R.~Petronzio and N.~Tantalo,
\newblock Phys.Lett. {\bf B595}, 408 (2004), [hep-lat/0405002].

\bibitem{Brandt:2013dua}
B.~B. Brandt, A.~J{\"u}ttner and H.~Wittig,
\newblock JHEP {\bf 1311}, 034 (2013), [1306.2916].

\bibitem{Bijnens:1998fm}
J.~Bijnens, G.~Colangelo and P.~Talavera,
\newblock JHEP {\bf 9805}, 014 (1998), [hep-ph/9805389].

\bibitem{PDG:2012}
Particle Data Group, J.~Beringer {\em et~al.},
\newblock Phys. Rev. {\bf D86}, 010001 (2012).

\end{thebibliography}

\end{document}